\documentclass[a4paper]{scrartcl}

\pdfoutput=1
\usepackage[english]{babel}
\usepackage{authblk}
\usepackage{booktabs}
\usepackage[utf8]{inputenc}
\usepackage[T1]{fontenc}
\usepackage{color}				
\usepackage{graphicx}			
\usepackage{amsmath}
\usepackage{amssymb,amsfonts,amsthm}
\usepackage{setspace}
\usepackage[english,hyperpageref]{backref}    
\usepackage{lipsum}
\usepackage{todonotes}
\usepackage{subfig}
\usepackage{hyperref,verbatim}
\usepackage{listings,dirtytalk}
\usepackage{lstlang3}
\usepackage{amsmath,amsfonts,amssymb,stmaryrd}
\usepackage{graphicx, wrapfig}
\usepackage{latexsym}
\usepackage{units}
\usepackage{xspace}

\usepackage{xcolor}
\hypersetup{
    colorlinks,
    linkcolor={red!50!black},
    citecolor={blue!50!black},
    urlcolor={blue!80!black}
}
\definecolor{bluekeywords}{rgb}{0.13,0.13,1}
\definecolor{greencomments}{rgb}{0,0.5,0}
\definecolor{redstrings}{rgb}{0.9,0,0}
\definecolor{codegreen}{rgb}{0,0.6,0}
\definecolor{codepurple}{rgb}{0.58,0,0.82}

\sbox0{\small\ttfamily A}
\edef\mybasewidth{\the\wd0 }
\usepackage{tikz}
\usetikzlibrary{arrows}
\usetikzlibrary{decorations.markings}
\usetikzlibrary{shadows}

\tikzset{
  big stealth/.style={
    decoration={markings,mark=at position -(0.1pt) with {\arrow[scale=2*\scale]{stealth}}},
    postaction={decorate},
    shorten >=0.4pt}}
\tikzset{
  big ring/.style={
    decoration={markings,mark=at position -(0.1pt) with {\arrow[scale=1.5*\scale]{o}}},
    postaction={decorate},
    shorten >=8pt*\scale}}
\tikzset{
  big disc/.style={
    decoration={markings,mark=at position -(0.1pt) with {\arrow[scale=1.5*\scale]{*}}},
    postaction={decorate},
    shorten >=8pt*\scale}}
\tikzset{
  big box/.style={
    decoration={markings,mark=at position -(0.1pt) with {\arrow[scale=1.5*\scale]{open square}}},
    postaction={decorate},
    shorten >=8pt*\scale}}
\tikzset{
  big tile/.style={
    decoration={markings,mark=at position -(0.1pt) with {\arrow[scale=1.5*\scale]{square}}},
    postaction={decorate},
    shorten >=8pt*\scale}}
\tikzstyle{place}=[circle, very thick, fill, top color=white, bottom color=white, draw=black, minimum size=40pt, drop shadow]
\tikzstyle{trans}=[rectangle, very thick, fill, top color=white, bottom color=white, draw=black, minimum size=32pt, drop shadow]
\tikzstyle{arc}=[thick, big stealth, black]
\tikzstyle{read}=[thick, big disc, black]
\tikzstyle{inhibitor}=[thick, big ring, black]
\tikzstyle{stopwatch}=[thick, big tile, black]
\tikzstyle{stopwatchinhibitor}=[thick, big box, black]
\tikzstyle{priority}=[thick, big stealth, orange]
\tikzstyle{enabling}=[thick, big disc, orange]
\tikzstyle{disabling}=[thick, big ring, orange]
\tikzstyle{token}=[circle, fill, draw=black, minimum size=4pt]
\tikzstyle{glob-options}=[label distance=6pt*\scalenodes*\scale,x=1pt,y=-1pt,scale=\scale,every node/.style={transform shape}]

\begin{document}

\title{Time-accurate Middleware for the Virtualization of
  Communication Protocols}

\author[1,2]{R.~Scarduelli\thanks{Les travaux décrits dans ce rapport
    ont éte soutenus par le projet de collaboration directe
    ``Représentativité temporelle des virtualisations de protocoles de
    communication'' avec la société Scalian, 2018.}}
\author[2]{P.-A.~Bourdil} \author[1]{S.~{Dal~Zilio}}
\author[1]{D.~{Le~Botlan}} \affil[1]{Université de Toulouse, CNRS,
  INSA, Toulouse, France} \affil[2]{SCALIAN, France}
\date{}
\maketitle
\begin{abstract}
  Communication between devices in avionics systems must be
  predictable and deterministic, and data must be delivered
  reliably. To help the system architects comply with these
  requirements, network protocol standards like ARINC 429 and AFDX
  were created. Even though the behaviour of each component in a
  network is well defined, it is still necessary to test extensively
  every applications before deployment. But physical test benches used
  in the aircraft development process are complex and expensive
  platforms. In order to limit the need for physical tests, we propose
  a time-accurate middleware for virtualizing communication protocols
  that can be used to replace physical tests with simulations.

  We specified three formal models of AFDX networks that take into
  account temporal constraints with different levels of precision. We
  also developed a prototype for a network virtualization middleware
  based on the AFDX protocol specification that provides an
  easy-to-setup environment for testing network
  configurations. Finally, we used formal models together with
  virtualization in order to define runtime monitors for checking
  whether the behavior of the middleware is time-accurate with respect
  to a real system.
\end{abstract}

\smallskip
\noindent \textbf{Keywords.}  Virtualization; avionics; AFDX; software
defined networking; Mininet; formal methods; communication protocols.


\newpage

\section{Introduction}

The growth in complexity of avionics systems---both for
flight-critical systems and for non-critical ones, such as passenger
entertainment---has fuelled an increase in the use of on-board
networks and data-buses. The desire for rapid deployment with minimal
development and implementation costs has driven the industry to
explore the choice of integrated, modular and off-the-shelf
technologies, such as the Avionic Full-Duplex Switched Ethernet (AFDX)
protocol, that supports deterministic data network buses targeting
aeronautical and military systems. Nonetheless, even tough the
behaviour of each component in a network (buses, routers, etc.) is
well defined, it is still necessary to test extensively every new
application before deploying it.

Signaling and inter-system communication in avionics have been a
crucial topic ever since electronic devices were first used in
aircraft. As time progressed, more and more systems which produce and
consume data were introduced in avionics, at some point becoming
crucial for even the most essential tasks, such as steering and later
fly-by-wire. To deal with these challenges in commercial avionics,
standards like ARINC 429~\cite{martinec_buckwalter_2014} were drafted
and adopted collectively by almost the entire industry. As the amount
of capabilities of aircraft operations are increasing, so is the
amount of information that needs to be processed and displayed. The
industry progressively requires a flexible and scalable standard
architecture to support the broad spectrum of capabilities and
performances. Consequently, the idea of \emph{Integrated Modular
  Avionics} (IMA)~\cite{morgan} was created, which introduces a number
of advantages over the traditional solutions, as resources now can be
shared and computational power can be added to the system when
necessary. As an evolved standard, ARINC 429 had many limitations, and
it could not cope with the ever increasing bandwidth, more flexible
topologies and new challenges like IMA, but it is still a proven and
commonly used protocol. The solution to the new challenge is to use
commercially proven hardware base technology and apply a protocol to
it, hence the \emph{Avionics Full-Duplex Switched Ethernet}
(AFDX)~\cite{afdxwhitepaper} protocol was specified---initially
developed by Airbus and later transformed into the actual ARINC 664
(Part 7) standard.

In avionics, communication between devices must be predictable and
deterministic, and data must be delivered reliably. Checking these
properties is usually obtained by testing every application
extensively on hardware that mimic the on-board network as faithfully
as possible. Physical test benches used in the aircraft development
process are complex platforms, with high initial and recurring
costs. They are generally on the critical path of the development and
cannot be easily multiplied to increase the validation
capacity. Hardware is one of the most expensive costs associated with
testing avionics network. This is why manufacturers try, as much as
possible, to use simulation or virtualization instead of physical test
benches. In this context,
\emph{virtualization}~\cite{network_virtualization} has many
advantages. First of all, it cuts out the need for deploying lots of
physical switches, wires, etc.; not only saving the costs of
maintaining and replacing those items, but also eliminating the costs
of powering those devices and paying for repairs. Virtualization also
makes development, testing, and deployment a lot faster. This is also
true when compared with simulation, since simulation is often much
slower than hardware, whereas virtualization can often run much
rapidly than actual avionics hardware. In addition, it is often easy
to make a virtualized network interact with real physical ``end
points'', or actual on-board software, in a kind of
\emph{Hardware-In-the-Loop} (HIL) approach.

This research report describes the work performed during a short
project between Scalian and the Vertics team at LAAS-CNRS. The purpose
of this project was to build and evaluate a \emph{time-accurate}
middleware for virtualizing communication protocols. Time-accuracy
meaning that the timing information obtained during a virtual
execution should be as faithful as possible to those observed in a
real system. The first objective is to develop and enrich a new
virtualization framework for AFDX based on the Mininet network
emulator and to analyze its timing behaviour. In our experiments, we
are mainly interested in computing the \emph{latency} and
\emph{jitter} values observed during a sequence of message
communications. Latency measures the time needed for a packet to
travel through the network, while {jitter} measures the delay
(deviation) between the actual and expected date of an event, like
sending a packet for instance.

At the same time, another goal is to evaluate the possible benefits of
using formal models with virtualization. The idea, here, is to define
formal models of AFDX networks that can precisely take into account
the temporal constraints of a real network. Then we can use components
of the formal models as reference points, or indicators, for checking
at runtime whether the behaviour of the middleware is compatible with
a network specification. In this work, we use the Fiacre specification
language---a formal language based on the theory of Time Petri
nets---to define the formal models and we use the Hippo execution
engine to transform a formal specification into a \emph{runtime
  monitor}.

We describe the main objectives pursued during the project. Along with
a brief description of the technologies used during this work, we
discusses the formal specification of an AFDX network with increasing
levels of approximation, corresponding to the precision of the models
with respect to the real system's operation. Thereby, a model for the
traffic policing mechanism present on such networks is also
described. Additionally, the development and implementation of the
virtualization middleware prototype is outlined, together with the
integration of real-time components.

The remainder of this document is organized as follows: Sect. 2
outlines some of the related works to this project. Sect. 3 gives an
overview of the main theory and technologies used and referred to
along the work. Sect. 4 explains the core concepts of the avionics
network protocol studied on this work and presents a use case. Sect.
5 details the formal specification of three models of an AFDX
network. Sect. 6 covers the development and implementation of
middleware for network virtualization. 

\subsection{Context of this Work}

This work stems form a research collaboration between Scalian, a
French company specializing in digital systems, and the VERTICS
research group (Verification of Time Critical Systems) at LAAS, the
CNRS Laboratory for Analysis and Architecture of Systems.


\section{Related Work}

This work is mainly related to the problem of simulating a network
protocol; in this particular case AFDX. The most novel part of our
approach is to favour virtualization over pure software
simulation. Another distinguishing fact is that we concentrate on the
``timing accuracy'' of the simulation, a concern that is rarely taken
into account in practice. There exists some related work, that we
briefly list thereafter.

Working Group 97 (WG-97) at EUROCAE (\url{https://www.eurocae.net/})
is currently defining the VISTAS protocol, a standard of virtual
simulation for tests of avionics systems in virtual bench networks,
based on previous research efforts (see for example the work of
Fourcade \textit{et
  al.}~\cite{fourcade_thebault_cloury_gaudaire_mattos_2013}). It
includes the definition of a protocol for the virtualization of
avionics protocols such as AFDX/A664 and A429. VISTAS operates at OSI
layers L3/L4 and relies on IP multicast to form the
point-to-multipoint communications. In our work, we experiment with
the Openflow protocol L2 concept of \textit{flows}, instead of IP
multicast, to implement frame forwarding. Also, at the time of
writing, VISTAS do not cover the early phases of development, where
having a virtualized network running on a standard PC configuration
could be very useful. Therefore our approach is mostly complementary
to the one targeted by VISTAS.

There are also several research projects that deal with similar
problem. Most of the work on AFDX is targeted towards the problem of
\emph{dimensioning} a network. That is how to deploy routers and
allocate ``virtual lines'' in order to respect the constraints of a
given workload. Nonetheless, some works are also concerned with
simulating existing networks. 

In their paper entitled \textit{"Deterministic OpenFlow: Performance
  Evaluation of SDN Hardware for Avionic
  Networks"}~\cite{deterministic_openflow}, the authors compare a
hardware implementation of Openflow (\textit{HP E3800} switch) with an
AFDX switch (Rockwell Collins AFDX-380). They focus on the performance
and determinism of a single Openflow switch in a representative, worst
case network configuration. While their focus is to add deterministic
behaviors to Openflow switches, we use them in the intent of
virtualization for test bench at early phases of development. The
authors conclude that the Openflow switch offers similar performances
as state of the art switch currently used in aircrafts for the most
parts of the network, and newer generation switches will likely be
able to handle the full network. Therefore, we believe that once a
virtual test bench is setup on a single PC it will be easy to add
determinism to the test bench by deploying it on hardware switches as
demonstrated by the authors.

We can also cite works that target other kind of network protocols,
like for example the Linux network stack.  Beifus \textit{et
  al.}~\cite{beifus_raumer_2015} study packet latency caused by the
packet processing software in PC systems based on Linux. The authors
created a simulation model using the Linux network stack, and
validated the model with respect to the packet latency based on
test-bed measurements with sub-microsecond accuracy. Their simulation
results showed the possibility to correctly predict the packet
latency, except for cases that occur outside of normal operating
parameters. We use some of the ideas from this study in order to
validate our model based on network parameters calculated from a real
network configuration, and extend the experiments to support low
latency packet processing (i.e. real-time support) as suggested by the
paper.

Finally, while our approach is mostly based on virtualization, we can
also cite works based on the use of \emph{discrete event simulators},
such as OMNeT++ (\url{https://omnetpp.org/}). This is a software-based
approach that is widely used to create simulation models for network
protocols. Our work mainly differs from discrete event simulators in
that protocols are executed in real time instead of being
discretized. In addition, we use the virtual Ethernet interfaces of
Linux which are seen exactly the same as physical Ethernet devices
from the application software, which provides a more realistic
behavior and a smooth path to running on real hardware.


\section{Technical Background}

This chapter describes the main concepts, technologies and tools that were used in this work, as well as in the research and development field. The theoretical background was the first stage of the project development, which allowed a better understanding of real-time communication protocols and formal modeling. Furthermore, a study about the available development tools and environments was also necessary. The following sections provide a brief overview of these topics and pointers to the bibliographic sources.

\subsection{Formal Methods}

Formal methods are system design techniques that use rigorously specified mathematical models to build software and hardware systems. In contrast to other design systems, formal methods use mathematical proof as a complement to system testing in order to ensure correct behavior.

Among these techniques, Petri nets~\cite{murata_1989} are a graphical and mathematical modeling tool suitable for describing concurrent, distributed and parallel systems. As an intuitively appealing graphical form of presentation, Petri nets are the model of choice on modeling network communication protocols.

\subsubsection{Petri Nets}

A Petri net~\cite{murata_1989} is a particular kind of directed graph, together with an initial state called the \textit{initial marking} $M_{0}$. The underlying graph \textit{N} of a Petri net is a directed, weighted graph consisting of two kinds of nodes, called \textit{places} (\textit{p}) and \textit{transitions} (\textit{t}), where arcs are either from a place to a transition or from a transition to a place. Places can, but not necessarily will, indicate a net's state (or marking), where transitions dictate how the state flows.  When a transition is fired, \textit{tokens} are removed from the place of origin and placed in the place of destination. In addition, arcs may have weights which determine how many tokens are removed from a place and how many are placed in the other.

\begin{figure}[ht]
\centering
\includegraphics[width=0.7\textwidth]{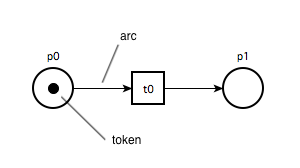}
\caption{\label{fig:petri-net-example}A simple Petri net example. When the transition \textit{t0} is fired, the token from place \textit{p0} flows to place \textit{p1}.}
\end{figure}

In a graphical representation, places are drawn as circles, transitions as bars or boxes, and tokens by \textit{k} black dots inside the places, where \textit{k} is the number of tokens available in that place. An example illustration of a simple Petri net is shown in Figure \ref{fig:petri-net-example}. A formal definition of a Petri net is given in Table \ref{tab:petri-net-definition}.

\begin{table}[ht]
\centering
\caption{Formal definition of a Petri net}
\label{tab:petri-net-definition}
\begin{tabular}{l}
\hline
A Petri net is a 5-tuple, $PN = (P, T, F, W, M_{0})$ where: \\
\\
\enspace $P = \{p_{1}, p_{2}, ..., p_{m}\}$ is a finite set of places, \\
\enspace $T = \{t_{1}, t_{2}, ..., t_{n}$ is a finite set of transitions, \\
\enspace$F \subseteq (P \times T) \cup (T \times P)$ is a set of arcs (flow relation), \\
\enspace$W: F \rightarrow {1, 2, 3, ...}$ is a weight function, \\
\enspace$M_{0}: P \rightarrow \{0, 1, 2, 3, ...\}$ is the initial marking, \\
\enspace$P \cap T = \varnothing$ and $P \cup T \neq \varnothing$. \\
\\
A Petri net structure $N = (P, T, F, W)$ without any specific initial marking is denoted by N. \\
\\
A Petri net with the given initial marking is denoted by $(N, M_{0})$\\
\hline
\end{tabular}
\end{table}

\subsubsection*{Extensions}

On top of the basic concepts of Petri nets, a number of extensions exist in order to model the systems more faithfully. The main extensions used on this documents are summarized below.

\begin{enumerate}
\item \textbf{Read arcs}: a read arc does not remove tokens from the place when the transition is fired.
\item \textbf{Prioritized Transitions}: adds priority to the transitions, where a transition cannot fire if a higher-priority transition is enabled (i.e. can fire).
\item \textbf{Time Petri Net}: enrich Petri nets with time intervals associated with the transitions of the net specifying the possible time delays between last enabledness of these transitions and their activation (or firing in Petri net terminology).
\end{enumerate}

\begin{figure}[ht]
\centering
\includegraphics[width=1\textwidth]{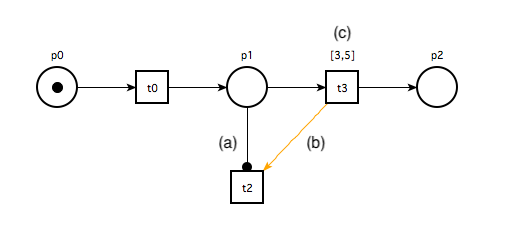}
\caption{\label{fig:petri-net-extensions}Petri net with mentioned extensions: (a) tokens from $p_{1}$ will not be removed when firing transition $t_{2}$; (b) transition $t_{2}$ cannot be fired if $t_{3}$ is enabled ($t_{3}$ has higher priority over $t_{2}$); (c) transition $t_{3}$ can only be fired within the time interval [3, 5] time units after it became enabled.}
\end{figure}

A Petri net using all these extensions is shown in Figure \ref{fig:petri-net-extensions}.

\subsubsection{Fiacre and TINA} \label{sec:fiacre-tina}

Fiacre~\cite{berthomieu:inria-00262442} is a specification language for describing compositionally both the behavioral and timing aspects of embedded and distributed systems. It has a formal semantics and can be used as an input format for formal verification tools (mainly real-time model-checkers) as well as for simulation purposes.

Fiacre stems from several projects involving industrial and academics partners, and is developed at LAAS. Besides the application described in this document, Fiacre has been used in a variety of applicative domains, like telecoms, avionics and robotics systems. In this work, we use Fiacre specifications with the model-checking toolbox TINA.

\subsubsection*{Fiacre Language}

Fiacre programs are structured into \textit{processes}, modeling sequential activities, and \textit{components}, describing a system as a composition of processes or other components. Fiacre supports the two most common coordination paradigms: by shared variable and by asynchronous message-passing.

\textit{Processes} are defined from a set of parameters and control states, each associated with a set of symbolic \textit{transitions} (following keyword \texttt{from}). The initial state is the source state of the first transition. The transitions declare how variables are updated, which events may occur, and when. They are built from standard deterministic programming constructs, non-deterministic constructs (such as external choice, operator \texttt{select}), communication statements, temporal constraints (construction \texttt{wait}) and jumps to a state (keyword \texttt{to} or \texttt{loop}).

\textit{Components} are built as parallel composition of processes and/or other components (by operator \texttt{par} $P_{0} || ... || P_{n}$ \texttt{end}). Compositions specify both the processes or component instances and their interactions. Shared variables and communication ports are within components. Communication ports may be associated with time constraints, applying to all interactions though those ports and with priorities. The ability to express timing constraints in programs is a distinguishing feature of Fiacre.

Introductory material and examples can be found on the Fiacre website\footnote{www.laas.fr/fiacre}. Also, some of the Fiacre specifications developed in this work are described later in Chapter \ref{cha:formal-models}.

\subsubsection*{TINA Toolbox}

Tina (TIme Petri Net Analyzer)~\cite{Berthomieu:2006:TPN:1173695.1173967} is a software environment to edit and analyze enriched Time Petri Nets. The core of the Toolbox is an exploration engine generating state space abstractions; these abstractions are then fed to model-checking or equivalence checking tools. The front-end converts models into an internal representation — Time Transition Systems (TTS) — an extension of Time Petri Nets with data and priorities. A compiler, \textit{frac}, converts Fiacre description into TTS descriptions, therefore enabling model-checking of Fiacre specifications by Tina.

\subsection{Software-Defined Networking}

Software-defined networking (SDN)~\cite{benzekki_2016} is an emerging networking paradigm. In an SDN, the network's control logic (control plane) is separated from the underlying routers and switches that forward the traffic (data plane). With the separation between the control plane and the data plane, network switches become simply forwarding devices and the control logic is implemented in a logically centralized controller (or network operating system), simplifying policy enforcement and network (re)configuration and evolution.

In this work we use the network prototyping software Mininet along with the OpenFlow protocol to virtualize avionics networks, more specifically AFDX.

\subsubsection{OpenFlow}

OpenFlow~\cite{openflow} is a communication protocol that controls a switch’s forwarding behavior. It is running between a controller and multiple switches. Packets can be matched on different fields (e.g. destination MAC address) and then associated to an action or a set of actions. Actions include forwarding to ports, but also rewriting of certain parts of the packet. OpenFlow defines a rule for each flow; if a packet matches a rule, the corresponding actions are performed (e.g. drop, forward, modify, or enqueue).

\subsubsection{Mininet} \label{sec:mininet}

Mininet~\cite{lantz_heller_mckeown_2010} is a system for rapidly prototyping large networks on the constrained resources of a single laptop. By combining lightweight virtualization with an extensible CLI and API, Mininet provides a rapid prototyping workflow to create, interact with, customize and share a software-defined network, as well as a smooth path to running on real hardware.

Users can implement a new network feature or entirely new architecture, test it on large topologies with application traffic, and then deploy the exact same code and test scripts into a real production network. Mininet runs surprisingly well on a single laptop by leveraging Linux features to launch networks with gigabits of bandwidth and hundreds of nodes (switches, hosts, and controllers).

Mininet uses the lightweight virtualization mechanisms built into the Linux OS: processes running in network namespaces, and virtual Ethernet pairs. Below are described the main components used by Mininet to create virtual networks.

\textbf{Links:} A virtual Ethernet pair, or veth pair, acts like a wire connecting two virtual interfaces; packets sent through one interface are delivered to the other, and each interface appears as a fully functional Ethernet port to all system and application software. Veth pairs may be attached to virtual switches such as a software OpenFlow switch.

\textbf{Hosts:} A host in Mininet is simply a shell process (e.g. bash) moved into its own network namespace. Each host has its own virtual Ethernet interface(s) and a pipe to a parent Mininet process, which sends commands and monitors output. Network namespaces are containers for network state. They provide processes with exclusive ownership of interfaces, ports, and routing tables. 

\textbf{Switches:} Software OpenFlow switches provide the same packet delivery semantics that would be provided by a hardware switch.

\textbf{Controllers:} The strategic point in SDN, controllers connect and configure the network devices (routers, switches, etc.). Controllers can be anywhere on the real or simulated network, as long as the machine on which the switches are running has IP-level connectivity to the controller.

\begin{figure}[ht]
\centering
\includegraphics[width=0.8\textwidth]{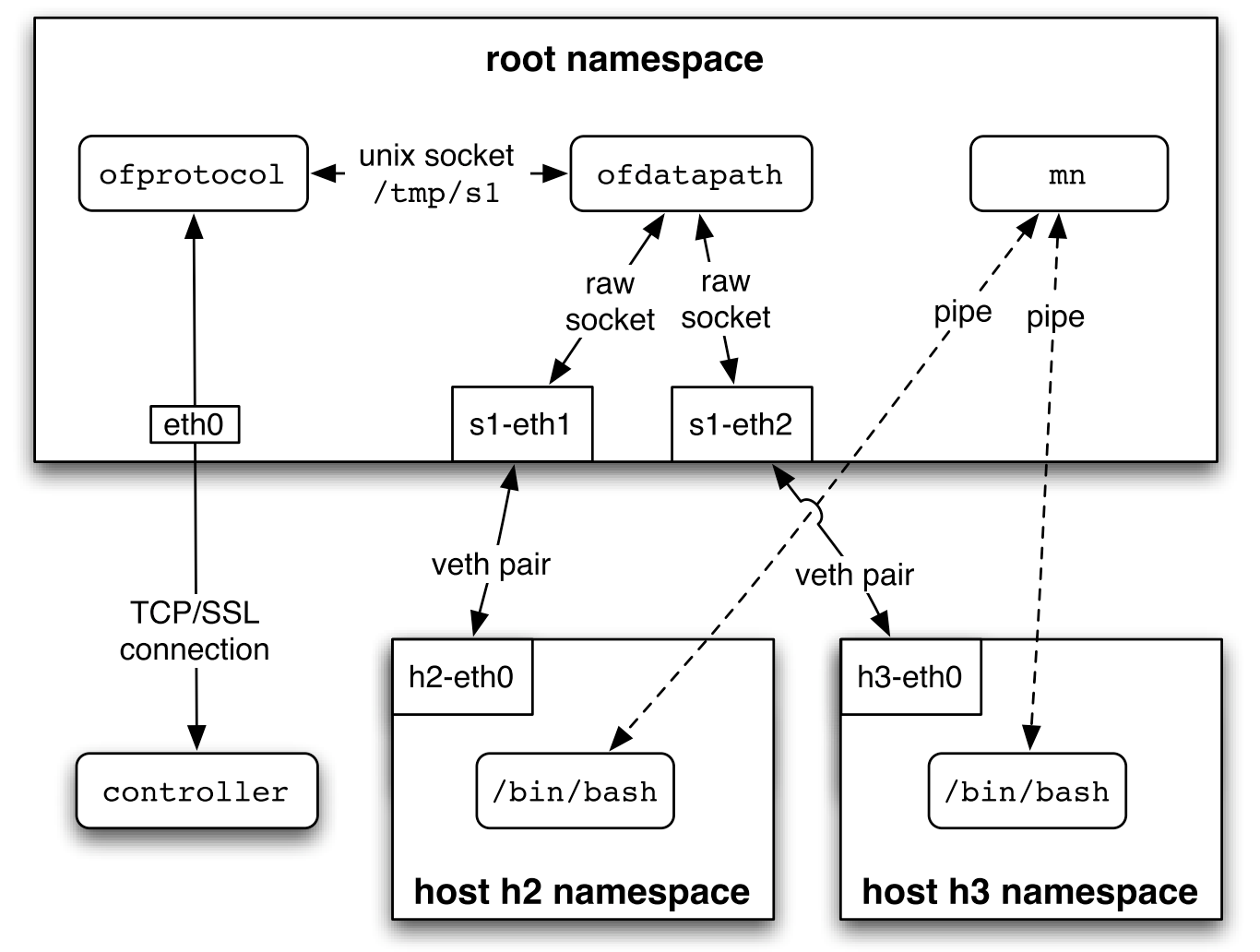}
\caption{\label{fig:mininet-example}Two-host virtual network example created with Mininet. Mininet creates a virtual network by placing host processes in network namespaces and connecting them with virtual Ethernet (veth) pairs. In this example, they connect to a user-space OpenFlow switch. Source: Lantz \textit{et al.}~\cite{lantz_heller_mckeown_2010}}
\end{figure}

Figure \ref{fig:mininet-example} illustrates the components and connections in a two-host network created with Mininet.

Mininet exports a Python API to create custom experiments, topologies, and node types (switch, controller, host, or other), which is used to create all the virtualized networks described later in this document.

\subsection{Real-time Operating Systems}

When operating with real-time operating systems (RTOS), reaching the desired result within the deadline is as important as reaching the result in the correct fashion. That is to say that tasks are required to be completed in time otherwise they lose their validity. 

For the purpose of ensuring deadline meeting, operating systems that run in real-time are available in many different forms. Since critical systems such as avionics networks rely on determinism, the use of a real time OS with this characteristic (\textbf{hard} real-time OS) is the choice for this project. Hard real-time OSes are systems that can deterministically meet a deadline, meaning that a task is always finished within its time-frame and being unable to do so is considered a failure.

\subsubsection{Xenomai}

Xenomai\footnote{https://xenomai.org/} is a hard real-time development framework that cooperates with the Linux kernel, providing real-time support to user space applications and at the same time seamless integration to the environment.

To support hard real-time capabilities to the Linux kernel, Xenomai implements a micro-kernel between the hardware and the Linux kernel. This micro-kernel is responsible for executing hard real time tasks and intercepts interrupts, blocking them from reaching the Linux kernel, hence preventing the Linux kernel from preempting the hard real-time micro-kernel.

Xenomai shows immense flexibility and ease of use for being Linux based, and is already being used in several projects at LAAS, hence the choice to use this real-time OS in this work.


\section{Overview of the AFDX specification}

This chapter describes the network protocol studied in this
project. The fundamental components of the protocol are described,
followed by a use case network which will be used to identify, clarify
and organize the system requirements for the next chapters.

\subsection{AFDX / ARINC 664}

AFDX is an acronym for Avionic Full-Duplex Switched
Ethernet~\cite{afdxwhitepaper}, a data network for safety-critical
applications that utilizes dedicated bandwidth while providing
deterministic quality of service. The network is based on standard
IEEE 802.3 Ethernet technology. 

There are many benefits from using commercial-off-the-shelf (COTS)
Ethernet components. This include reduced overall costs, faster system
development and easier maintenance. However, standard commercial grade
Ethernet do not meet avionics network requirements. For this reason,
AFDX was defined as an extension of Ethernet standard that adds
support for Quality of Service (QoS) and deterministic behaviour.

An AFDX network consists of so called End Systems and switches, where
point-to-point or point-to-multipoint connections are represented by
Virtual Links (VL). We explain the main concepts of AFDX networks in
this chapter.

\subsection{End Systems}

An End System (ES) is a component connected to the AFDX network
capable of handling all AFDX related protocol operations. Usually, an
End System is part of an avionic or aircraft subsystem, which needs to
send or receive data over the AFDX network.

This interface guarantees a secure and reliable data interchange with
other avionics subsystems. One or more switches, depending on the
network hierarchy, are located on the data path between two End
Systems.

\begin{figure}[ht]
\centering
\includegraphics[width=0.9\textwidth]{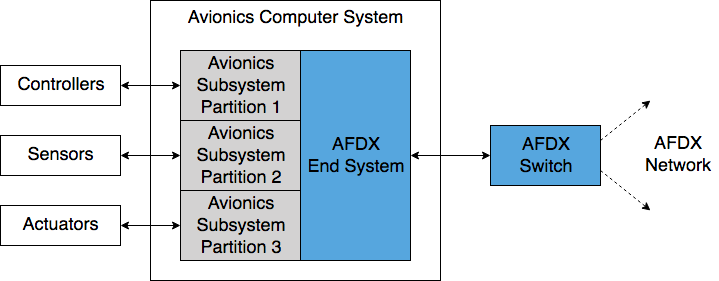}
\caption{\label{fig:afdx-end-system}Example of an End System connected
  to different avionics subsystems.}
\end{figure}

As shown in Fig.~\ref{fig:afdx-end-system}, an avionics computer
system connects to the AFDX network through an End System. In general,
an avionics computer system is capable of supporting multiple avionics
subsystems, isolated from one another through partitioning. The End
System will then multiplex the traffic from the avionics subsystems
onto one single wire, and it must be capable of doing so respecting
several timing constraints, some of which are discussed later in this
document.

\subsection{Virtual Links}

One of the precursors to AFDX is ARINC 429, a standard developed for
safety-critical applications in 1977 (30 years before AFDX was
patented). One of the most desirable features of an ARINC 429
connection~\cite{martinec_buckwalter_2014} is the fact that it
represents a private line between sender and recipient(s) of
data. Hence the physical bandwidth of the connection is available at
all times, and no concurrent transmit requests can occur. This degree
of separation is mandatory when interconnecting avionics systems with
different levels of criticality. On the minus side, the ARINC 429
approach means that every connection needs a dedicated pair of wires,
which can be difficult to lay out on a plane and can add significant
weight. AFDX achieves an equivalent result by choosing logical
connections, referred as \emph{Virtual Links} (VL), over physical
ones. VL are based on partitioning the time at which communication can
occur on a physical link.

\begin{figure}[ht]
\centering
\includegraphics[width=1\textwidth]{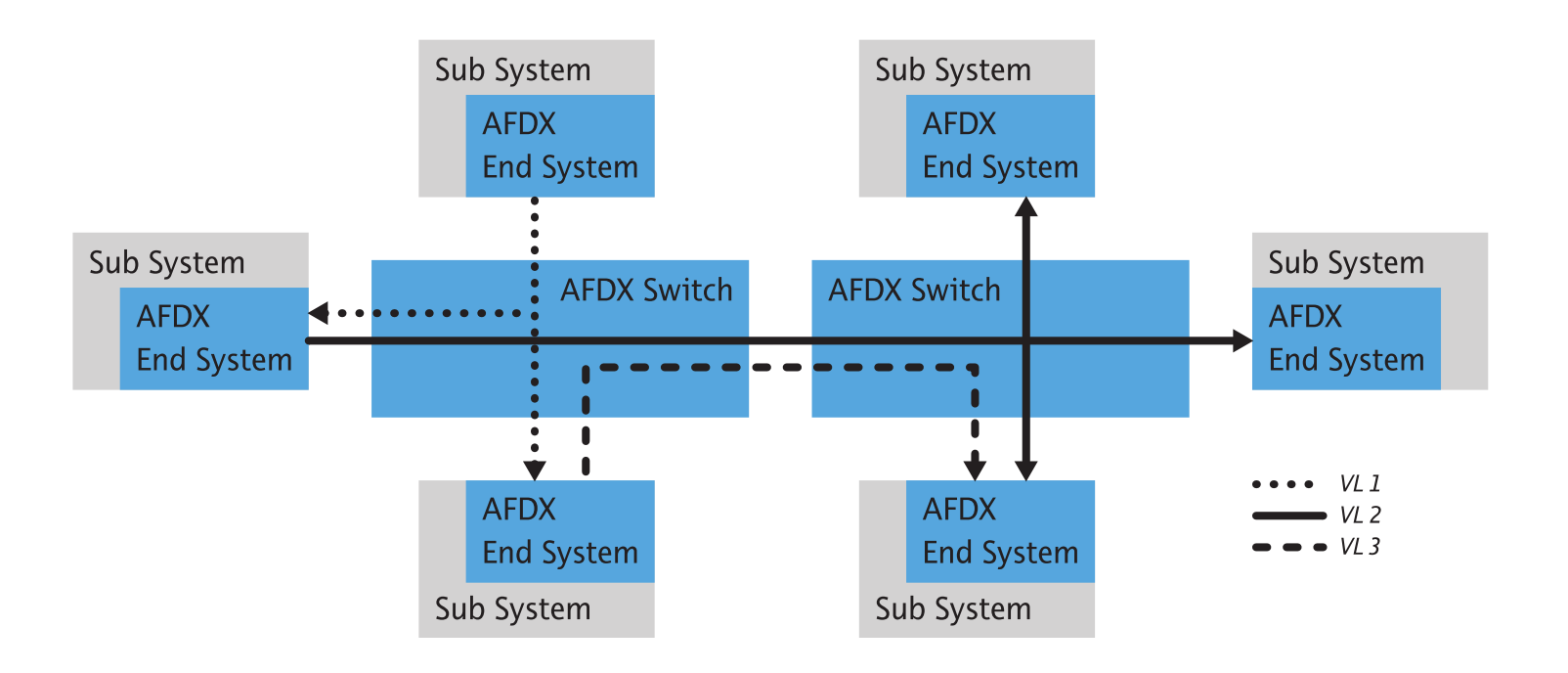}
\caption{\label{fig:afdx-virtual-links}Virtual Link topology. Source: AFDX White paper~\cite{afdxwhitepaper}}
\end{figure}

In an AFDX network, each logical connection is represented by a
Virtual Link. Each VL builds a unidirectional logic between a unique
source ES to one or more target ES, thus providing a private line with
bounded latency and guaranteed bandwidth. An example of Virtual Link
topology is given Fig.~\ref{fig:afdx-virtual-links}, where the network
has three VL, connecting six End Systems using only two switches.

\subsection{Switch}

The core of an AFDX network is built using switches, which have more
responsibilities than their counterpart used in commercial Ethernet
network. In addition to the obvious switching functions, an AFDX
switch must perform frame filtering and traffic policing duties,
ensuring that traffic arriving at the switch is compliant with the
restrictions set for the appropriate VL.

\begin{figure}[ht]
\centering
\includegraphics[width=1\textwidth]{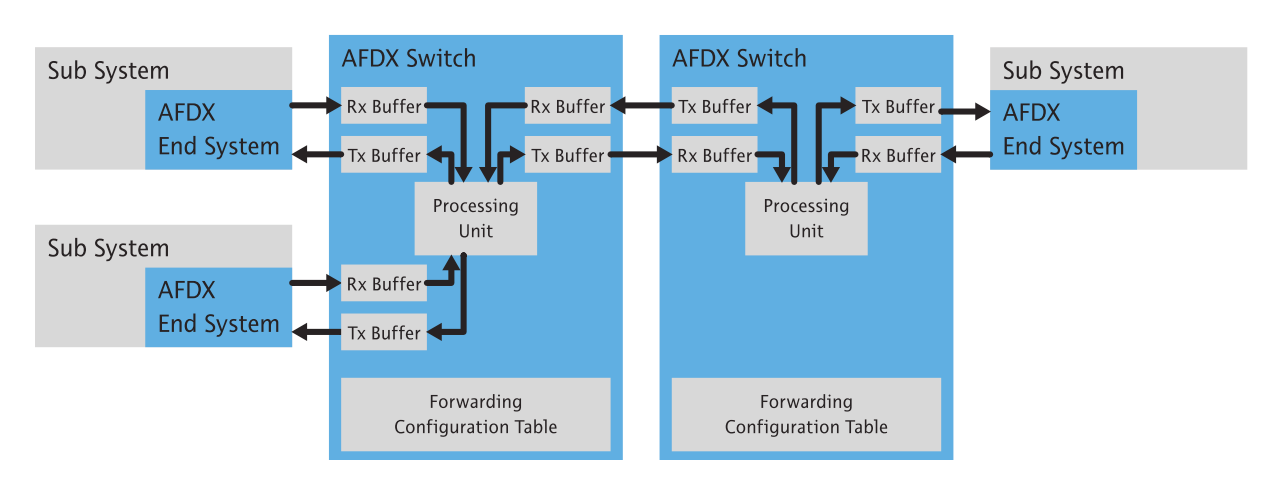}
\caption{\label{fig:afdx-switch}AFDX Subsystem to Switch communication. Source: AFDX White paper~\cite{afdxwhitepaper}}
\end{figure}

In order to decouple transmit operation from data reception, all data
paths use separate data buffers, thus creating true full duplex data
links between End Systems. Because updating the data paths while the
switch is operating introduces variable latency, which is not
acceptable for avionic data networks, an AFDX switch forwards packets
according to a static configuration table. The main components of an
AFDX switch is shown in Figure \ref{fig:afdx-switch}.

\subsubsection{Traffic Policing}

The AFDX standard specifies a traffic policing function at the switch
based upon the token-bucket algorithm common to switched-packet
networks. The goal of traffic policing is to ensure that no VL exceeds
its allotted bandwidth. This algorithm will be explained in depth in
the next Chapter.

\subsection{Guaranteed Service}

The AFDX protocol is oriented towards ensuring guaranteed service at
every level; both the bandwidth and maximum end-to-end latency of the
link are guaranteed. However, there is no guarantee of packet
delivery. Packets may be dropped, for example, if they will exceed the
limit imposed by traffic policing. Since AFDX is only concerned by the
transport layer, transmission acknowledgements and re-transmissions
requests must be handled at the application level.

\subsubsection{BAG} \label{sec:bag}

\begin{figure}[ht]
\centering
\includegraphics[width=0.45\textwidth]{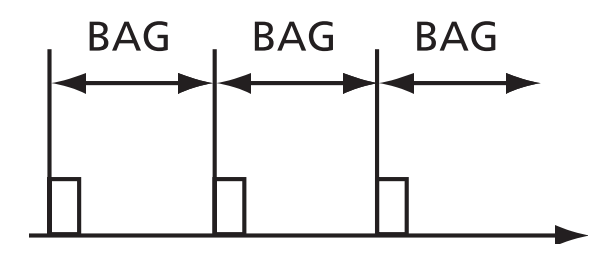}
\caption{\label{fig:bag}Bandwidth Allocation Gap. Source: Developing AFDX Solutions~\cite{actel}}
\end{figure}

The primary bandwidth control mechanism is the BAG (Bandwidth
Allocation Gap) (Figure \ref{fig:bag}). The BAG defines the minimum
time interval between the starting bits of two successive AFDX frames,
assuming zero jitter. The AFDX specification allows for BAG values
that are in the range of 1..128 ms and that are a power of $2$; that
is a bag is always of the form $2^{n}$ with $n \in \{0,\dots,7\}$.

Each VL has an associated BAG and a maximum frame size, but a VL may
not use all the bandwidth, i.e. the time between two consecutive
packets may be larger than BAG. However, it may not be smaller than
BAG and it is the responsibility of the ES to enforce this.

\subsubsection{Jitter} \label{sec:jitter}

\begin{figure}[ht]
\centering
\includegraphics[width=0.9\textwidth]{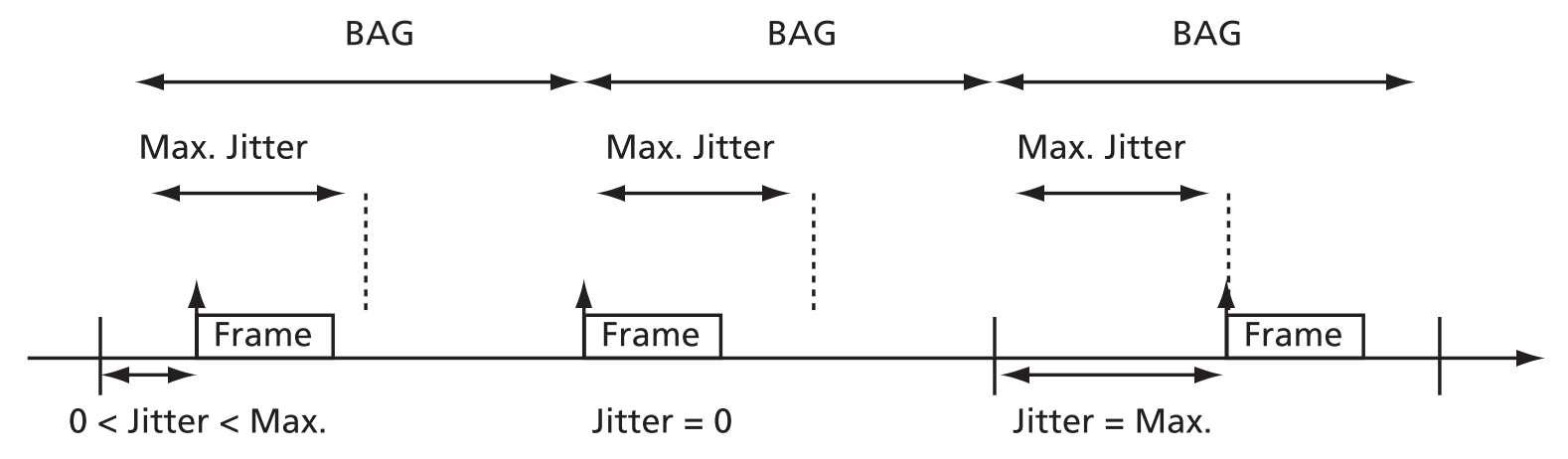}
\caption{\label{fig:jitter}Jitter Defined. Source: Developing AFDX Solutions~\cite{actel}}
\end{figure}

The ES may introduce jitter when transmitting frames for a given
VL. This jitter is defined as the delay between the beginning of the
BAG and the date when the first bit of the frame is sent (Figure
\ref{fig:jitter}).


A given ES may have to transmit data for multiple VLs, so a frame from
one VL can be delayed up to the maximum allowed jitter value to limit
the instantaneous ES frame rate and thus accommodate frames from other
VLs. The maximum allowed jitter for a given ES is defined by
equation~\ref{eq1}, below:

\begin{equation} \label{eq1} \text{Max. Jitter} \leq 40\mu s +
  \frac{\sum\limits_{j \in \{ VLs\}} ((20 + Smax_{j}) *
    8)}{\text{Nbw}}
\end{equation}

\begin{equation} \label{eq2}
 \text{Max. Jitter} \leq 500\mu s
\end{equation}

\noindent where $\text{Nbw}$ is the link bandwidth (100 Mbps), based
on standard IEEE 802.3 10/100 Mbit Ethernet hardware and equal for all
VLs, and $S_{max}$ is the maximum allowed frame size for the VL (in
number of bytes). As seen from equation~\ref{eq1}, the specification
allows a minimum of 40 $\mu s$ for the ``technological'' jitter, while
the ``second part'' part of the formula takes into account the time
needed to transfer the frames payload. In no case the total jitter can
exceed a hard-limit of 500 $\mu s$, see equation~\ref{eq2}.

\subsubsection{Latency}
Latency measures the time needed for a packet to
travel through the network.

Although AFDX does not specify a maximum system latency, any supplier
is required to specify the upper limit of latency for any system
delivered.

\subsection{Frame Format}

\begin{figure}[ht]
\centering
\includegraphics[width=1\textwidth]{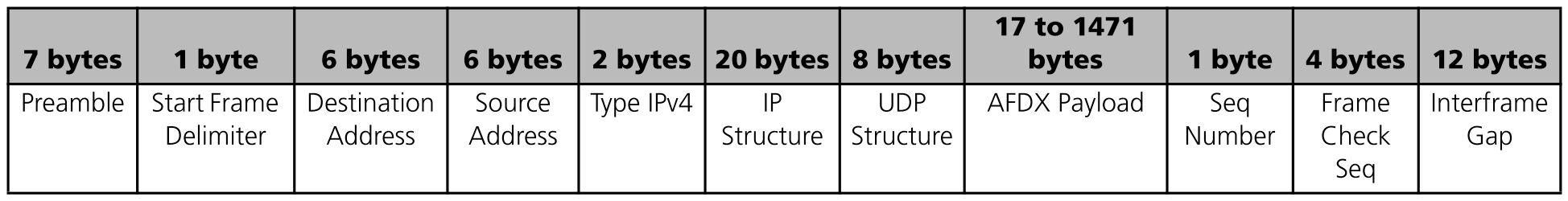}
\caption{\label{fig:afdx-frame}AFDX Frame. Source: Developing AFDX Solutions~\cite{actel}}
\end{figure}

The AFDX frame format is shown in Figure \ref{fig:afdx-frame}. 

The one-byte sequence number is used to maintain ordinal integrity
within a given VL. The frame sequence number is initially set to 0
upon ES start-up or reset. During continuous operation, the number
wraps back to 1 after reaching a value of 255.

The maximum frame size is set for each VL and is represented by the
parameter $S_{max}$. The range of this parameter is between 64 and
1518 bytes.

The destination and source addresses listed contain the MAC addresses
for the ESes. AFDX network addressing is based upon the MAC addresses,
which are 16-bit in length. The source address must be a unicast
address and follow the format detailed in the specification, and
includes bits for identifying to which of the two redundant networks
the MAC is attached. The destination address is a multicast address
that includes a 16-bit VL identifier.

\subsection{Redundancy}

An AFDX network is constructed so there are two independent paths
(including MACs, PHYs, and cabling) between each ES, as well as
redundant switches to protect the network from a failure at the MAC
level or below. The default mode is to transmit the same frame (with
identical frame sequence numbers) across both networks, but the
redundancy option can be configured so that frames for a given VL may
be sent along either or both of the networks.

The redundancy management is done at application level. The receiving
ES then accepts the first valid frame and passes it to the
application. Once a valid frame is received, any other frame with the
same sequence number is discarded.

\subsection{Use Case: Flight Management System} \label{sec:use-case}

We have used several example of AFDX networks to benchmark our
approach. While most of them are generated randomly, we have also used
a representative example of network extracted form the PhD thesis of
Michael Lauer~\cite{lauer}, which is a subset of the navigation
system, called Flight Management System (FMS), of an Airbus plane. The
purpose of this system is to control the display of navigation
information on the flight screens used by the pilots. We detail this
network since it gives a good idea of the typical complexity of an
``avionic function''. A plane usually operates a thousand such
functions.

\begin{figure}[ht]
\centering
\includegraphics[width=0.8\textwidth]{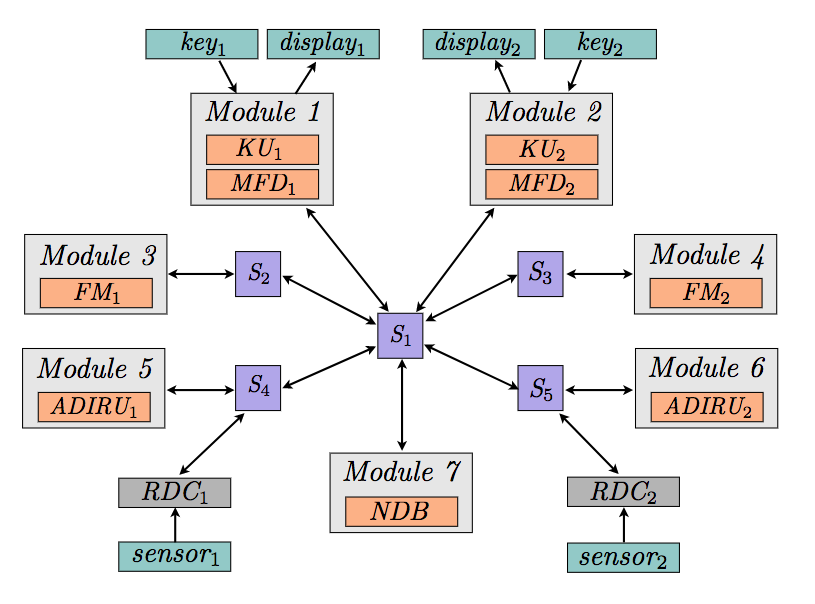}
\caption{\label{fig:fms-vl}Flight Management System network architecture. Source: Lauer~\cite{lauer}}
\end{figure}

The architecture of the FMS is composed of a set of modules
interconnected through an AFDX network. This architecture is shown in
Fig.~\ref{fig:fms-vl}. Seven modules, from Module 1 to Module 7, are
used to host the avionics functions. The Remote Data Concentrators
($RDC_{1,2}$) connect the pressure sensors ($sensor_{1,2}$) to the
network. Each of these modules contain an End System responsible for
handling all AFDX related protocol operations. The keyboards
($key_{1,2}$) and displays ($display_{1,2}$) however are connected
directly to the modules via a field-bus. The AFDX network consists of
five switches, S1 to S5, represented by the purple boxes in the
figure.

Table \ref{table:vl} shows the configuration for each VL of the network.

\begin{table}[ht]
\centering
\caption{FMS Virtual Links.}
\label{table:vl}
\begin{tabular}{@{}lllll@{}}
\toprule
VL   & source      & destination      & BAG (ms)  & $S_{max}$ (bytes)\\ \midrule
$VL_{1}$  & $KU_{1}$    & $FM_{1}, FM_{2}$ & 32        & 75\\
$VL_{2}$  & $KU_{2}$    & $FM_{1}, FM_{2}$ & 32        & 75\\
$VL_{3}$  & $FM_{1}$    & $MFD_{1}$        & 8         & 625\\
$VL_{4}$  & $FM_{1}$    & $NDB$            & 16        & 125\\
$VL_{5}$  & $FM_{2}$    & $MFD_{2}$        & 8         & 625\\
$VL_{6}$  & $FM_{2}$    & $NDB$            & 16        & 125\\
$VL_{7}$  & $NDB$       & $FM_{1}$         & 64        & 500\\
$VL_{8}$  & $NDB$       & $FM_{2}$         & 64        & 500\\
$VL_{9}$  & $RDC_{1}$   & $ADIRU_{1}$      & 32        & 64\\
$VL_{10}$ & $RDC_{2}$   & $ADIRU_{2}$      & 32        & 64\\
$VL_{11}$ & $ADIRU_{1}$ & $FM_{1}, FM_{2}$ & 32        & 87.5\\
$VL_{12}$ & $ADIRU_{2}$ & $FM_{2}, FM_{1}$ & 32        & 87.5
\end{tabular}
\end{table}

This network will be used extensively during this document as basis
for modeling, virtualization, simulation and benchmarking. However,
this topology is only a small part of a full network system and is
considered to be small to industry standards. Indeed, a typical
industrial AFDX network can have more than 100 ES, two redundant
sub-networks with 8 switches each, with thousands of VLs (more than
6000 paths) on each sub-network. In order to test the scalability of
our tools, we will often use variations of this use-case where case
network topology and parameters are tweaked when applicable to
represent different, more realistic scenarios.

\subsection{Benchmarking} \label{sec:benchmarking}

I have developed two generators of AFDX networks (specifications) in order to validate and to benchmark our virtualization framework. The first generator is used to build a simple AFDX network with $N$ Virtual Links, such that: all the VLs have only one source and one destination; the bag and the frame size of each VL is chosen randomly; and all the VLs go through a single (shared) router. The idea is to generate a new, random network topology by varying the following three parameters:
\begin{itemize}
    \item Number of Virtual Links: this is the main factor that drives the complexity of the network. The VLs are generated as point-to-point links connected to a single switch.
    \item BAGs: for each VL the emission period is chosen from the set of available BAGs \{1, 2, 4, 8, 16, 32, 64, 128\}$ms$.
    \item Frame size: frames emitted on the VLs can be of any size between 64 and 1518 \emph{bytes}. For each VL, the frame size is constant throughout the simulation
\end{itemize}

In addition to randomly generating the topology, some simulation parameters can also be arbitrarily chosen to cover a bigger spectrum of network behavior:

\begin{itemize}
    \item Speed factor: slows down or speeds up the network emission periods. For example, a speed factor of 2 will make all the VLs send frames twice as fast.
    \item Duration: the simulation duration, where we consider short to be 10 seconds, medium 60 seconds (1 minute) and long 300 seconds (5 minutes). Since the emission period is in the order of milliseconds, even the shorter duration produces a considerable amount of traffic.
\end{itemize}

I have also developed a second kind of generator based on templates. The idea is to start from a known network (expressed in a CSV file, see for example Listing~\ref{lst:csv-file}) and to generate bigger version by taking several copies of it. Network specifications can be saved using a textual formats and reused with the different tools that we developped during my internship.

By looking at the traffic over time, more specifically the jitter and latencies of each Virtual Link, a detailed analysis of the network can be accomplished. This information yields how the traffic over the network is shaped---who may be transmitting data at a higher rate than configured, or whose traffic is the most affected by the current configuration.

While studying the time series of the traffic provides a simple yet informative visualization of the network dynamics, they are not very practical for studying finer-scale features. Traces can become long enough that in small scales any plot is dominated by over-plotting, and does not provide much useful information. This is specially true when comparing two time series that are already rather similar. In this dissertation we make use of an alternative method called \emph{Cumulative Distribution Function} (CDF), which enables the analysis and comparison of traces at finer scales.

\begin{figure}[ht]
\centering
\includegraphics[width=1\textwidth]{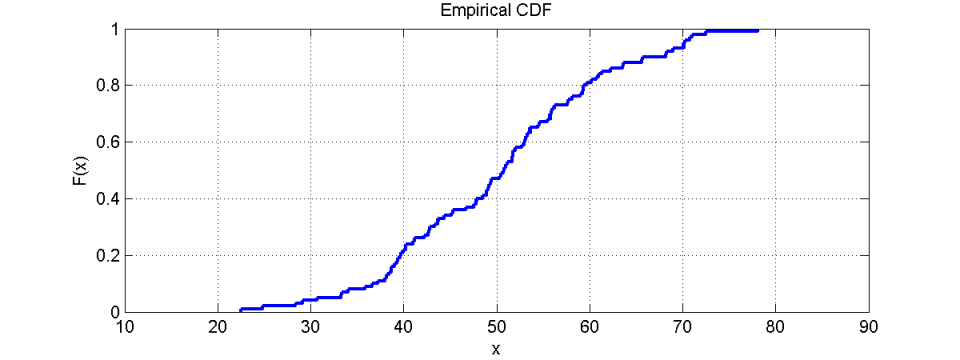}
\caption{\label{fig:empirical-cdf}Empirical CDF plot. Source: http://www.andata.at}
\end{figure}

Figure \ref{fig:empirical-cdf} illustrates an empirical example of a CDF plot. This plot means that a relative amount of $F(x)$ values from the given set of numbers used to calculate this distribution are less or equal than the value $x$.

This plot enables a direct quantitative reading of relevant key values. For example, the minimum can be seen right at the point where the CDF begins and hits the x-axis---i.e. zero probability of having values that are smaller than $F(x)$ when $x=0$. Similarly, the maximum can be seen where the CDF reaches the line $y=1$ and ends. Percentiles can also be read directly from the x-axis. The existence of outliers also becomes apparent, and it is showcased later through some experimental results.


\section{Three Formal Models of AFDX Networks} 
\label{cha:formal-models}

This section discusses the modeling of the main components of an AFDX
network using the formal specification language Fiacre (see
Section~\ref{sec:fiacre-tina}). We describe three different abstractions
that corresponds to different levels of precision on how frames are
transported across the network. In doing so, we also define a formal
model of the \emph{token bucket} traffic policing mechanism. We
conclude by giving some experimental results obtained from simulation
of these models. These formal models are the first contribution of this work.

\subsection{Modeling with the Fiacre language}

We model the behaviour of an AFDX network using three different
approaches: (1) timed channels; (2) direct virtual links; and (3)
switched virtual links. Instead of defining the complete encoding of a
whole network, we consider only the model of a single Virtual Link,
that is Virtual Link 1 ($VL_{1}$) of the FMS example given in
Sect.~\ref{sec:use-case}. Since Fiacre supports composition of
processes and components to create more complex models, we can easily
build the whole network from this simple example.

Virtual Link 1 is used to transport frames from the Keyboard Unit
partition $KU_{1}$ on Module 1, to both Flight Manager partitions
$FM_{1}$ and $FM_{2}$ on Modules 3 and 4, respectively. Since we
consider frames with a minimum-sized payload of $17$ bytes, we can
derive the following properties from Table \ref{table:vl} and
equation~\eqref{eq1}:

\begin{itemize}
    \item \textbf{BAG = 32 ms}. The minimum time between two consecutive frames.
    \item \textbf{J$_{max}$ = 47.6 $\mu$s}. The maximum amount of jitter introduced by the End System.
    \item \textbf{S$_{max}$ = 75 bytes}. The maximum frame size transmitted on this VL.
\end{itemize}

We focus our model on the network part of the system. Therefore we do
not model the behaviour of End Systems and simply assume that each VL
transmits with maximum bandwidth, meaning that every frame has a
constant size and is sent by the ES exactly at each BAG.


We start with the model of a simple periodic communication protocol,
see Listing~\ref{lst:sendrecv}. This example has two communicating
process instances (Sender and Receiver) encapsulated in a component
(System) which defines the communication port and the timing
constraints. For example, channel \texttt{p} is declared with a timing
interval of $[32, 32]$, meaning that every interaction
(synchronization) on this channel must be performed exactly 32 units
of time after the two processes are willing to communicate. 

In Fiacre, the expression \texttt{p!\,send} stands for an emission on
port \texttt{p}, carrying the value of variable
\texttt{send}. Symmetrically, the expression \texttt{p?\,x} stands for
a reception on port \texttt{p}; the value ``carried'' by the port is
assigned to variable \texttt{x}. We can view each of our encodings of
AFDX as a more refined example of this simple system.

\begin{lstlisting}[language=Fiacre,float=bt,captionpos=b,caption=\protect{Example of sender and receiver system on Fiacre.},label=lst:sendrecv,frame=single]
process Sender [pout : int] is
    states send
    var msg : int := 0
    from send
        pout! msg; to send
        
process Receiver [pin : int] is
    states recv
    var msg : int
    from recv
        pin? msg; to recv
        
component System is
    port p : int in [32, 32]
    par p in
       Sender   [p]
    ||  Receiver [p]
    end
    
\end{lstlisting}

\subsubsection{Timed Channel} \label{sec:timed-channel}

The first network approximation model is based on components called
\emph{timed channels}. The idea is that the whole ``travel'' of a
frame through the switches and data buffers of the actual network can
be abstracted away as a simple delay on a dedicated channel.  In the
encoding of a whole network, we simply associate one timed channel to
each path in a VL. Hence each channel represents a link between a
single sender and a single receiver.

A timed channel is characterized by a time interval $[a, b]$, where
$a$ and $b$ are the best (lowest) and worst network traversal time for
a frame in a given VL, also known as Worst Case Traversal Time (WCTT)
and Best Case Traversal Time (BCTT). The traversal time corresponds to
the sum of the technological latencies of the switches crossed along
the path and the output transmission time for each of these
devices. These bounds can be calculated using techniques such as
Network Calculus~\cite{LeBoudec:2001:NCT:1755809} or Trajectory
Approach~\cite{martin2004maitrise}.

\begin{figure}[ht]
\centering
\includegraphics[width=0.65\textwidth]{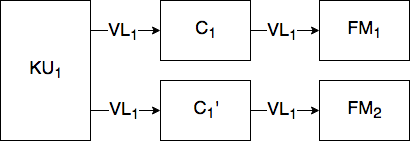}
\caption{\label{fig:vl1-timed-channel}Timed channels approximation of $VL_{1}$.}
\end{figure}

Figure \ref{fig:vl1-timed-channel} shows a diagram of how this
approximation is applied to Virtual Link 1. $VL_{1}$ is decomposed
into two timed channels, $C_1$ and $C'_1$, each one transmitting
frames every 32 ms (BAG) with a delay in the interval $[298, 444]$
(here the unit of time is the $\mu$s). The full use case network
architecture with timed channel approximation and the corresponding
parameters are given on Appendix~\ref{a1}.

Listing \ref{lst:delay} illustrates the timed channel process which
acts as the communication medium between the communicating
processes. The sender process writes a new frame into variable
\texttt{vin} every BAG ms, meaning that the value of \texttt{vin}
becomes different from \texttt{NIL}; then, as soon as this condition
is true (the Fiacre keyword \texttt{on} is used to define a guard on a
transition), process TimedChannel waits for a delay in the interval
$[298, 444]$ then copies the frame to \texttt{vout} and resets
\texttt{vin} to \texttt{NIL} so it can wait for the next frame;
finally the receiver process reads the frame from the \texttt{vout}
variable whenever there is a frame available.

\begin{lstlisting}[language=Fiacre,float=bt,captionpos=b,caption=\protect{Fiacre model of timed channels.},label=lst:delay,frame=single]
type Frame is NIL | VALUE end

process TimedChannel (&vin, &vout : Frame) is
    states get
    from get
        on (vin <> NIL);
        wait [298, 444];
        vout := vin;
        vin := NIL;
        to get
\end{lstlisting}

It is worth noting that network traversal times are always smaller
than the period at which frames are generated (the BAG). This holds
true for the other timed channels as well (see Table
\ref{table:tc}). This means that a timed channel will always delay
only one frame at a time, which greatly simplifies our model. Also,
this first implementation does not take into account the jitter at the
sender process. However, the following models takes into account
jitter and it would be easy to adapt the same approach in this case
also.

\subsubsection{Direct Virtual Link}

In our second approach, we take into account the multicast nature of
Virtual Links as if they were direct links between the sender and
receivers partitions, but without including the switches.

\begin{figure}[ht]
\centering
\includegraphics[width=0.6\textwidth]{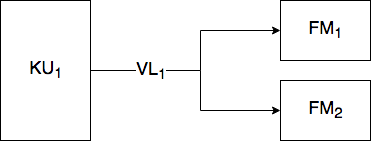}
\caption{\label{fig:vl1-virtual-link} Approximation of $VL_{1}$
  behavior without the switches.}
\end{figure}

Each VL is composed by a unique sender process (in contrast to one for
each timed channel), and one or more receiver processes. As shown in
Figure \ref{fig:vl1-virtual-link}, $VL_{1}$ originates from one port
at partition $KU_{1}$, and is synchronized to both destinations
$FM_{1}$ and $FM_{2}$.

We should not model this approach as is. Indeed, if we are not
careful, and since receivers are synchronized directly with the same
sender, both would receive the frames at the same time. This is
obviously not true in a real network since traversal time can vary on
different paths. One way to introduce different traversal time for
each path is to make the receiver (Rx) be a composition of two
different processes, here called \texttt{RxPort} and \texttt{RxDelay}
as shown by Listing \ref{lst:rx}. Process \texttt{RxPort} is
synchronized with the sender by the communication port \texttt{pin},
but we do not consider the reception of a frame complete by this
process. Instead, it sends the frame to \texttt{RxDelay} using a
shared variable, which will introduce a variable delay to only then
mark the frame as received.

\begin{lstlisting}[language=Fiacre,float=ht,captionpos=b,caption=\protect{Fiacre model of the VL receiver component (Rx).},label=lst:rx,frame=single]
type Frame is NIL | VALUE end

process RxPort [pin : Frame] (&frame : Frame) is
    states recv
    from recv
        pin? frame; to recv
        
process RxDelay (&frame : Frame) is
    states idle, delay
    from idle
        on (frame <> NIL); wait [0, 0]; to delay
    from delay
        wait [BCTT, WCTT]; frame := NIL; to idle    /* frame received */
        
component Rx [pin : Frame] is
    var frame : Frame := NIL
    par pin in
       RxPort [pin] (&frame)
    ||  RxDelay      (&frame)
    end
\end{lstlisting}

The Tx component has a similar mechanism to introduce jitter on frame
emission. It is a composition of a \texttt{TxBAG} process, which sends
frames through a shared variable to a \texttt{TxJitter} process, who
then introduces a bounded delay corresponding to the VL's maximum
jitter specification before sending the frame on the communication
port. Since the hard limit for the maximum admissible jitter is 500
$\mu$s (EQ. \ref{eq2}), which is smaller than the lowest BAG (1 ms)
(see Sect.~\ref{sec:bag}), we do not have to worry about
\texttt{TxJitter} having to delay more than one frame at a time.

\subsubsection{Switched Virtual Link}

Our last modelling approach is the closest one to the actual physical
behaviour of the network. In particular, it takes into account
switches and the routing of frames until they reach their destination.

\begin{figure}[ht]
\centering
\includegraphics[width=0.8\textwidth]{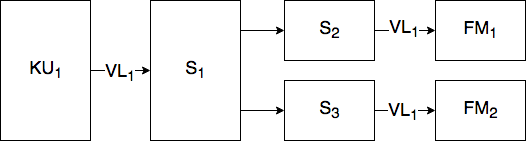}
\caption{\label{fig:vl1-switch}$VL_{1}$ with the switch model. The switches forward the frames from the sources to its destinations.}
\end{figure}

In this approach, each partition is synchronized to a switch through
input and output communication ports. The switches are configured with
a static routing table that will forward the frames to the correct
output ports. Figure \ref{fig:vl1-switch} shows this concept applied
to $VL_{1}$, where we see that each channel goes through two distinct
pair of switches.

Since we do not model the traffic shaping mechanism from the End
Systems, which combines different VLs into one single connection, we
still need separate ports for each VL originating on the same
partition.

This model provides the possibility to specify hardware constraints
and allows for a more detailed implementation of the AFDX
behaviors. For example, one of the most important is the traffic
policing mechanism present at the switch which provides guarantee on
the bandwidth for every VL.

Modeling the traffic policing contraption with formal methods adds
flexibility when integrating the network model with real or
virtualized systems. Because this algorithm has some properties that
are not trivial to model in Fiacre, such as maintaining a continuous
account, the next section is dedicated to it since it is a major part
of this work's contribution.

\subsection{Traffic Policing}

AFDX specifies a traffic policing function at the switch based on the
\emph{token bucket algorithm}, that is common to many packet-switch
networks. The goal of traffic policing is to ensure that no Virtual
Link exceeds its allotted bandwidth.

The specification requires the use of one or both of the following two
algorithms for traffic policing:

\begin{itemize}
\item Byte-based policing
\item Frame-based policing
\end{itemize}

As shown in~\cite{yao_qiu_kwak_2009}, the byte-based token bucket
algorithm may not be able to perform the traffic policing as
expected. Therefore, frame-based policing was chosen for this
implementation.

\subsubsection{Frame-based Token Bucket Algorithm}

The token bucket filter implements a resource-based algorithm to
decide whether to accept or reject a frame. The concept is quite
simple. A bucket has a current balance that cannot be negative. The
bucket is credited as time progress but cannot exceed a maximal
value. Each accepted frame has a cost, deducted from the bucket and a
frame is allowed only if the current balance is high enough. 

In an AFDX network, each VL has an independent bucket that is
controlled by the first switch found coming out of the source ES.  For
a given $VL_{i}$, an account $AC_{i}$ is maintained and credited with
tokens (in bytes) at the following rate:

\begin{equation} \label{eq1-tbf}
S_{i}^{max} / BAG_{i}
\end{equation}

\noindent where $S_{i}^{max}$ is the maximum allowed frame size and
$BAG_{i}$ is the Bandwidth Allocation Gap for the $VL_{i}$.  From the
AFDX standard, tokens accumulate in the account ($AC_{i}$) until it
reaches a limit value of $AC_{i}^{max}$, such that:

\begin{equation} \label{eq2-tbf}
AC_{i}^{max} = S_{i}^{max} * (1 + J_{i}^{max} / BAG_{i})
\end{equation}

\noindent where $J_{i}^{max}$ is the maximum defined jitter value for
VL number $i$.

\begin{figure}[ht]
\centering
\includegraphics[width=1\textwidth]{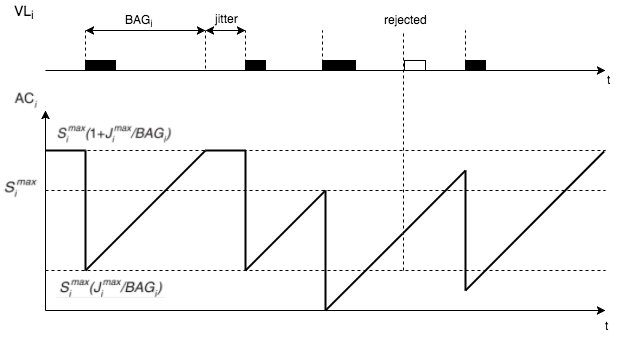}
\caption{\label{fig:tokenbucket}Example of frame-based traffic policing}
\end{figure}

Assume a frame of size $S_{i}$ is emitted on $VL_{i}$. When it arrives
at the first switch, it is allowed if the current account value,
$AC_{i}$, is greater than $S_{i}^{max}$. In this case $AC_{i}$ is
decremented by $S_{i}^{max}$. If the frame account $AC_{i}$ is less
than $S_{i}^{max}$, the frame is rejected and the value of $AC_{i}$ is
left unchanged. Figure \ref{fig:tokenbucket} illustrates this
behavior.

\subsubsection{TPN Model}

On the face of it, the behaviour of the token bucket seems difficult
to model. Indeed, the time at which a new frame is allowed may depend
on the time the previous frame was seen. Which indicates that we may
be faced with a system that has an hysteresis; an hybrid
system. Surprisingly, it is possible to model frame-based policing
using an ordinary Time Petri Net (TPN). The model if given in
Fig.~\ref{fig:tokenbucket}. The idea is to consider four possible
cases, each represented by a different state (or place in this case),
\texttt{s0} to \texttt{s3}.

\begin{figure}[ht]
\centering
\includegraphics[width=1\textwidth]{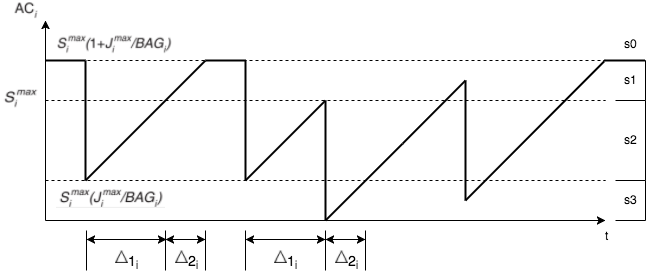}
\caption{\label{fig:tbf-tpn-graph}Frame-based policing parametrization to model as TPN}
\end{figure}

Place \texttt{s0} is used to model a state where the token bucket has
reached its maximal value, as shown in
Fig.~\ref{fig:tbf-tpn-graph}. This is also the initial state of the
system. Place \texttt{s1} corresponds to a state where a frame
arriving at the switch will be accepted. In between these two states,
we use places \texttt{s2} and \texttt{s3} to represent the transient
state where frame can be rejected. The bucket is associated with two
time constants, $\Delta_{1}$ and $\Delta_{2}$, that correspond,
respectively, to the time needed for the bucket to accept a frame
after ``leaving'' state \texttt{s0} and the time needed for reaching
its maximal balance. Place \texttt{s2} is used when a new frame can be
accepted less than $\Delta_1$ in the future. Place \texttt{s3} is used
when a frame is accepted before the bucket reached its maximal value.

As the account $AC_{i}$ for a Virtual Link $VL_{i}$ is credited at
rate $J_{i}^{max} / BAG_{i}$ (EQ. \ref{eq1-tbf}), the values of
$\Delta_{1i}$ and $\Delta_{2i}$ are defined by:

\begin{equation} \label{delta1}
\Delta_{1i} = (S_{i}^{max} - S_{i}^{max} * \frac{J_{i}^{max}}{BAG_{i}}) * \frac{BAG_{i}}{S_{i}^{max}}= BAG_{i} - J_{i}^{max}
\end{equation}

\begin{equation} \label{delta2}
\Delta_{2i} = [S_{i}^{max} * (1 + \frac{J_{i}^{max}}{BAG_{i}}) - S_{i}^{max}] * \frac{BAG_{i}}{S_{i}^{max}} = S_{i}^{max} * \frac{J_{i}^{max}}{BAG_{i}} * \frac{BAG_{i}}{S_{i}^{max}} = J_{i}^{max}
\end{equation}

\begin{figure}[ht]
\centering
\includegraphics[width=1\textwidth]{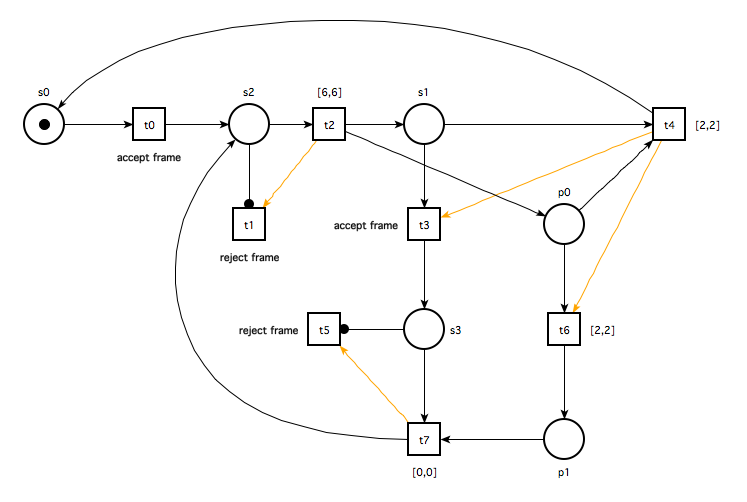}
\caption{\label{fig:tbf-tpn.png}Frame-based token bucket for VL with $BAG = 8$ and $J_{max} = 2$}
\end{figure}

Figure \ref{fig:tbf-tpn.png} shows an example implementation of the
frame-based token bucket policing applied to a VL with $BAG = 8$ and
$J_{max} = 2$ time units. These values were chosen in order to
simplify calculations and ease understating, but they do not represent
a real VL configuration.

Because we need to keep track on how much time has elapsed when accepting frames while on place \texttt{s1}---time that is necessary to transition back from \texttt{s3} to \texttt{s2}---we need to add two additional places: \texttt{p0} and \texttt{p1}. This behavior simulates the token bucket account increasing with an steady rate, where accepting frames leaves the account in a state proportional to how "filled" it was before.

Priorities here are necessary mainly because of the ambiguity when transitioning from accepting- to rejecting-states and vice versa. Since the frame accepting policy is based on comparing its size to how much tokens---normally represented by bytes---is available on the account, when the right amount of time has elapsed equivalent to the frame size, the algorithm dictates that it must be accepted. Thus we prioritize accepting frames over rejecting right at the state boundaries, when appropriate. Furthermore, owing to the way we model the keeping of time from state \texttt{s1}, the two rightmost priorities ensure the proper transition to state \texttt{s0} when the account is full.

\subsubsection{Fiacre implementation}

\begin{figure}[ht]
\centering
\includegraphics[width=1\textwidth]{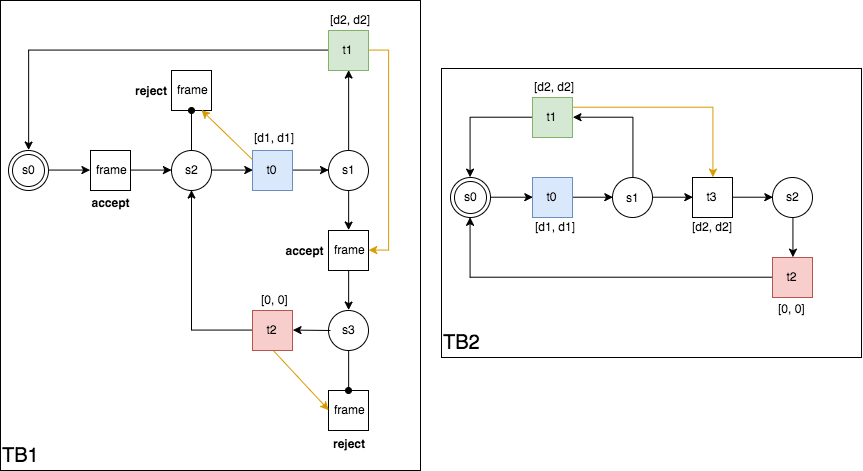}
\caption{\label{fig:tbf-fiacre.png}Fiacre processes that compose the frame-based token bucket TPN model.}
\end{figure}

We can define a Fiacre specification that is equivalent to the TPN
given in the previous section. To implement this TPN model in Fiacre,
a composition of two processes is necessary. Figure
\ref{fig:tbf-fiacre.png} display the two processes, \texttt{TB1} and \texttt{TB2} , which are synchronized in transitions \texttt{t0}, \texttt{t1} and \texttt{t2} to create the full model \texttt{TokenBucket}. Transition times \texttt{d1} and \texttt{d2} correspond to $\Delta_{1}$ and $\Delta_{2}$ respectively.

The full Fiacre code to implement this model is listed on Listing \ref{lst:tbf-fiacre}. This code does not have means of directly integrating with external components (e.g. sender and receiver), since the communication port \texttt{frame} is local to the \texttt{TokenBucket} component and it is only used to flag the arrival of frames, but it showcases the algorithm mechanism and gives a full, neat example on how a TPN is translated to the Fiacre specification language.

\begin{lstlisting}[language=Fiacre,float=ht,captionpos=b,caption=\protect{Token bucket algorithm in Fiacre.},label=lst:tbf-fiacre,frame=single]
process TB1 [t0, t1, t2, frame : sync] is
    states s0, s1, s2, s3
    from s0 
        frame; to s2              // accept frame
    from s2
        select 
            frame; loop           // reject frame
        []  t0;  to s1 
        end
    from s1
        select 
            frame; to s3;         // accept frame
        []  t1;  to s0 
        end
    from  s3
        select 
            frame; loop           // reject frame
        []  t2;  to s2 
        end

process TB2 [t0, t1, t2, t3 : sync] is
    states s0, s1, s2
    from s0
        t0; to s1
    from s1
        select 
             t1; to s0
        []   t3; to s2 
        end
    from s2
        t2; to s0

component TokenBucket is
    port
        t0      : sync in [d1, d1],   // d1 = BAG - Jmax
        t1, t3  : sync in [d2, d2],   // d2 = Jmax
        t2      : sync in [0, 0],
        frame   : sync                // frame reception
    priority
        t1 > t3,
        t0 | t1 | t2 | t3 > frame
    par * in
        TB1 [t0, t1, t2, frame]
    ||   TB2 [t0, t1, t2, t3]
    end
\end{lstlisting}

\subsection{Experimental Results}

We developed a tool to generate traces from the simulation of formal models. It is based on the stepper simulator \emph{play}, which comes with the TINA Toolbox and allows the simulation step by step of net descriptions in several formats, including Time Transition Systems (TTS). Our tool loads the net and randomly advances the time, firing transitions in no particular order as they become enabled. To generate the traces, we can specify the transitions of interest in order to obtain the date at which they are fired.

\begin{figure}[ht]
\centering
\includegraphics[width=1\textwidth]{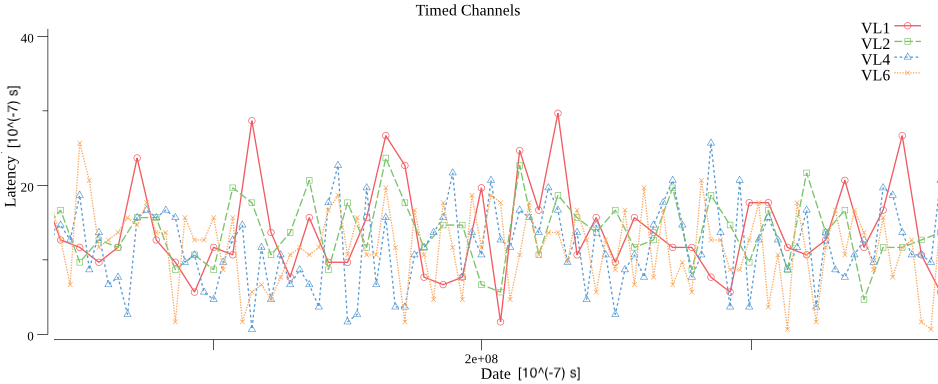}
\caption{\label{fig:formal-vl-plot}Latency trace for four VLs during simulation from formal models.}
\end{figure}

While all levels of abstraction can be simulated, this report only present results from the most complex one: switched virtual links. I modeled the whole use case network---complete with all switches, End Systems and traffic policing---using the concepts described in this section. Figure~\ref{fig:formal-vl-plot} shows the latency of four VLs obtained from a 60-seconds simulation of the whole network. 

\begin{figure}[ht]
\centering
\includegraphics[width=1\textwidth]{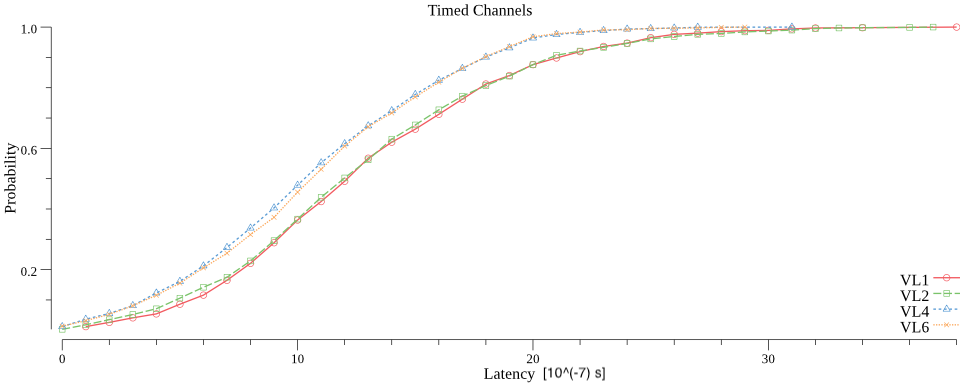}
\caption{\label{fig:formal-cdf-plot}Latency CDF for VLs simulated from formal models.}
\end{figure}

Figure~\ref{fig:formal-cdf-plot} illustrates the latency CDF calculated from this simulation. We can see from these graphs a similarity between Virtual Links due to the redundancy present on AFDX configurations---$VL_{1}$ and $VL_{2}$ have the same parameters and cross the same number of components when traversing the network.

The simulation to real time ratio achieved by this tool is 1:1, meaning that our simulation runs at the same speed as a real network. Latency is introduced in these simulations by changing model parameters such as the switch's output processing time, chosen arbitrarily in this case to be 0.1 $\mu$s at maximum.

\section{Virtualization of AFDX Network}
\label{cha:netw-virt}

This chapter details the development and implementation of the
prototype middleware for our network virtualization tool. Firstly, the
main components developed to work with the Mininet framework are
described, followed by how we implement the traffic policing mechanism
on the Linux platform. Secondly, several experimental results are
analyzed, and a strategy to circumvent some of the problems
encountered is outlined.

\subsection{AFDX Virtualization with Mininet}

We already introduced Mininet in Section~\ref{sec:mininet}. Mininet
creates a virtual network by placing host processes in network
namespaces and connecting them through \emph{virtual Ethernet} (veth)
pairs connected, for example, to a user-space OpenFlow switch.

In this chapter, we use the concepts learned during the modelling of
AFDX networks and the different approximations that we used for our
formal models to create a virtual AFDX network based on Mininet. We
analyse the viability of this approach and show ways to improve the
simulation of a network.

\subsubsection{Hosts}

We use the concept of \emph{Hosts} to represent End Systems in
AFDX. Since this work focuses on implementing the network behaviours,
we simplified some of the mechanisms on the application level: hosts
represent the end-points of a Virtual Link; for a given VL with one
source and $K$ destinations there is $K+1$ (Mininet) hosts: one for
the source---responsible for sending frames---and one for each target
ES---the receiver hosts.

\subsubsection*{Sender Hosts}

In each of the sender hosts, throughout the simulation, a task sends
frames through its \textit{veth} interface periodically with respect
to the VL's BAG in order to generate its average bandwidth, as
illustrated in Listing \ref{lst:send}.

The periodic tasks are simulated using a high resolution timer with
Linux soft real time scheduling, and the tasks are given the highest
execution priority ($99$).

Due to a limitation of virtual Ethernet devices and Precision Time
Protocol (PTP) Linux driver, timestamping at emission is done in the
host node just before writing the frame into the socket.

\begin{lstlisting}[language=C,float=bt,captionpos=b,%
caption=\protect{C code of the task running at sender host nodes (simplified).},%
label=lst:send,frame=single]
/* Set high priority */
/* Configure socket */
/* Build AFDX frame */

for (int i = 0; i < COUNT; i++) {       // send COUNT frames periodically
    add_seqno(&frame);                  // add sequence number to frame
    clock_gettime(CLOCK_REALTIME, &ts); // get timestamp
    sendto(sockfd, frame, ...);         // send frame on socket
    nanosleep(BAG * 1000000);           // wait BAG milliseconds
}

/* Save timestamps to file */
\end{lstlisting}

\subsubsection*{Receiver Hosts}

Similar to the sender hosts, each VL destination is represented by a
task which is listening to frames arriving at its \textit{veth}
interface throughout the simulation, as shown in Listing
\ref{lst:recv}.

At reception, frames are (software) timestamped using the PTP Linux
driver, which is more precise than the way timestamping is done at
emission. This difference in timestamping mechanisms may compromise
the data when calculating network metrics such as latency, but is
still sufficiently precise to serve its purpose.

\begin{lstlisting}[language=C,float=h,captionpos=b,caption=\protect{C code of the task running at receiving host nodes (simplified).},label=lst:recv,frame=single]
/* Configure socket */

for (int i = 0; i < COUNT; i++) {       // receives COUNT frames
    recvmsg(sockfd, &frame);            // receive frame on socket
    /* Apply system macros for accessing the ancillary data */
    memcpy(&ts, CMSG_DATA(cmsg), ...);  // get software timestamping from frame
}

/* Save timestamps to file */
\end{lstlisting}

\subsubsection{Switch}

At the OpenFlow switch where all the hosts are connected through
virtual Ethernet pairs, we use the concept of \textit{flows} to
implement frame forwarding.

The matching of VLs is based on the frame's destination MAC
address. OpenFlow allows for exact and wildcard matching of
destination MAC addresses. To separate the flows from each other, an
exact matching entry for each MAC address is needed. To allow for
AFDX's multicast behavior, multiple actions can be associated to each
entry, e.g. by having multiple output actions.

This configuration must be done prior to beginning the network
simulation. Openflow allows for installation of rules during the
runtime of a switch. However, such behavior is not desired in
avionics, especially not during flight. Therefore, we assume the
installation of rules is only performed during a maintenance phase on
ground and then leave all configuration as-is until the next
maintenance.

Listing \ref{lst:switch} illustrates how the \textit{flows} are
configured on the OpenFlow switch using the Mininet Python API. Each
flow contains the input port in which the Virtual Link enters the
switch, the Ethernet destination address to match the rule (Virtual
Link ID), and the outputs to forward the matched VL's packets to. With
all the flows defined, we use the \texttt{dpctl} utility tool to add
them to the flow table.

\begin{lstlisting}[language=Python,float=h,captionpos=b,caption=\protect{OpenFlow switch configuration with Mininet Python API},label=lst:switch,frame=single]
# list containing all flows
self.flows = [{"in_port": 1,                # input port
               "vl_id":   "00:01",          # virtual link ID
                actions:  "2,3"}, ... ]     # output ports

# forwarding table rules
fwd_afdx_port = "\"table=0
                 in_port=%d 
                 dl_dst=03:00:00:00:%s/ff:ff:ff:ff:ff:ff
                 actions =%s\""

# add forwarding rules for each flow
for flow in self.flows:
    self.dpctl("add-flow", fwd_afdx_port % (flow["in_port"], 
                                            flow["vl_id"], 
                                            flow["actions"]))
\end{lstlisting}

\subsubsection*{Traffic Policing}

To implement the traffic policing on the switch we use linux traffic
control (tc). Traffic control is part of the linux iproute2 package
which allows the user to access networking features.

With tc it is possible to attach queuing disciplines (qdiscs) to
network interfaces. Queueing disciplines are packet queues with an
algorithm that decides what to do with ongoing or incoming packets. In
every network interface there is an \textit{ingress} qdisc which works
on incoming traffic. Therefore, in order to classify and police frames
entering on a switch veth interface corresponding to the VL parameters,
one need to add a token bucket algorithm to the \textit{ingress}
qdisc---this can by done by adding filters with a police action
attached to it.

The filter provides a convenient mechanism for gluing together several
of the key elements of traffic control. The simplest and most obvious
role of the filter is to classify packets. One of the most common
classifying mechanisms is the u32 classifier. The u32 classifier
allows the user to select packets based on attributes of the
packet. In order to classify differently for each VL, we use the u32
classifier to match according to the destination MAC address of the
AFDX Ethernet frame, which contains the VL identifier.

The police action allows to limit bandwidth of traffic matched by the
filter it is attached to. From the algorithms available to measure the
packet rate, we use the token bucket which is configured using the
rate, burst, overhead and conform-exceed parameters.

\begin{itemize}
\item \textbf{rate}: The maximum traffic rate of packets passing this
  action. Those exceeding it will be treated as defined by the
  conform-exceed option. This can be calculated for each VL by
  $S_{i}^{max} / BAG_{i}$ (EQ \ref{eq1-tbf}).
\item \textbf{burst}: Set the maximum allowed burst in bytes. For
  $VL_{i}$, the burst parameters sets the amount o acceptable jitter,
  and can be calculated by $S_{max} * (1 + J_{i}^{max} / BAG_{i})$.
\item \textbf{overhead}: Account for protocol overhead of
  encapsulating output devices when computing rate. For Ethernet
  frames, the overhead parameter is 14 bytes.
\item \textbf{conform-exceed}: Define how to handle packets which
  exceed the configured bandwidth limit. This is configured to
  \textit{drop} in order to be compliant with the AFDX specification.
\end{itemize}

Below is an example command to add traffic control to a virtual
Ethernet interface from the switch (s1-eth0) receiving frames from
$VL_{1}$ (03:00:00:00:\textbf{00:01}), which has parameters BAG = 32
ms, S$_{max}$ = 75 bytes and J$_{max}$ = 0.5 ms.

\begin{verbatim}
  $ tc qdisc add dev s1-eth0 ingress
  $ tc filter add dev s1-eth0 parent ffff: \
        match ether dest 03:00:00:00:00:01 \
        police rate 2344bps burst 77b overhead 14 conform-exceed drop
\end{verbatim}
      
\subsubsection{Topology}

\begin{figure}
\centering
\includegraphics[width=1\textwidth]{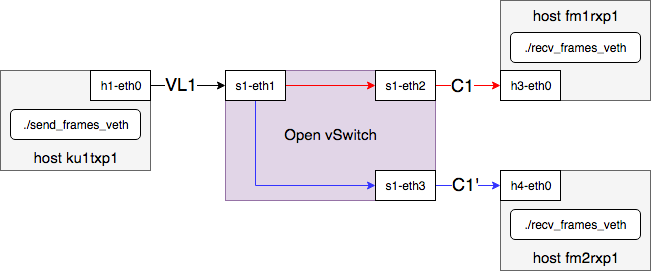}
\caption{\label{fig:mininet-vl1}Many-to-one mapping applied to $VL_{1}$}
\end{figure}

To create the topology of the AFDX network we apply a many-to-one
mapping to the real topology, that is we use a single switch to
configure the path of all VLs. This can be thought as implementing the
Virtual Link approximation presented on
Chapter~\ref{sec:timed-channel} with a single intermediate switch to
perform traffic policing and create the point-to-multipoint
connections for all the VLs, so we can characterize each path as a
timed channel. Figure \ref{fig:mininet-vl1} illustrates this mapping
applied to $VL_{1}$.

\subsection{Monitors}

To gather information about the virtualized network performance, we
have implemented three different monitor: one for checking whether the
observed latency is out of bounds (with respect to the AFDX
specification); a second for checking the value of jitter at emission
(to detect if the sending site does not respect its realtime
constraints); and finally a monitor to compute the number of frames
dropped due to the traffic policing mechanism. Monitors implement a
kind of runtime verification that could allow us to tag the frames
that are outside the normal behaviour of the simulated network (and
hence that should be disregarded in the simulation).

The timestamps referred in this document correspond to Unix time with
nanoseconds resolution, and are computed with a different technique at
emission and reception sites, as discussed in the previous section.

\subsubsection{Latency}

The latency monitor is used to verify if the network traversal time
for frames on each VL path is conform with the Best (resp. Worst) Case
Traversal Time (see Table \ref{table:tc}).

In order to calculate the latency on-line (during the network
simulation), the timestamp at emission is encapsulated by the payload
before writing the frame into the socket. Therefore, when a frame is
received by a destination process, the reception timestamp is then
subtracted from the emission timestamp parsed from the payload to
obtain the latency.

If the latency is greater than the WCTT for the corresponding path
(timed channel), the frame is flagged and saved to a buffer, which
will be dumped into a log after the end of simulation for further
analysis. If the latency is smaller than the BCTT it means that we can
run the simulation faster than the real network.

\subsubsection{Jitter} \label{sec:jitter-monitor}

When generating frames in the sender processes, the period BAG is not
strictly respected, since the tasks need processor time which is
shared among the other processes running on the machine. Consequently,
a technological jitter is introduced to the BAG.

To calculate the jitter, the emission timestamp for each frame is
compared with the theoretical emission time if the BAG was always
strictly respected.

Say the jitter of the $i$-th frame is calculated. We classify the
jitter according to four categories:

\begin{enumerate}
\item \textbf{Too early}: the frame was emitted before its normal
  period, that is before date $i * BAG$.
\item \textbf{OK}: the frame was emitted in the acceptable interval
  $[i * BAG, i * BAG + J_{max}]$.
\item \textbf{Too late}: the frame was emitted after the maximum
  admissible jitter $J_{max}$ but it did not missed the period (it is
  sent before date $(i+1) * BAG$).
\item \textbf{Skipped Period}: The frame has missed its period; that
  is it was emitted more than $BAG$ ms after date $i * BAG$.
\end{enumerate}


\subsubsection{Traffic Policing}

To check how many frames are being dropped by the traffic policing
mechanism, we use the sequence number present on the AFDX frame.

Each frame has a one-byte sequence number that increments with every
frame sent. From discontinuities on the sequence number parsed at
reception, we can infer the number of frames lost in transmission.

\subsection{Experimental Results}

The experiments are done on a Virtual Machine running Linux (1 CPU, 2048MB Memory) on top of a standard MacBook Pro laptop (Intel Core i5, 8GB Memory), using Mininet. 

By analyzing the network traffic on multiple network configurations, mainly link latency and emission jitter, a number of problems emerged and modifications on the code and experimental setup became necessary to eliminate or mitigate the 
most relevant ones.

The rest of this chapter describes the virtualization framework evolution, explaining the reasoning behind solving the main obstacles encountered during development. For brevity, I only display the results related to the use case network configuration, but the benchmarking tools described in Section~\ref{sec:benchmarking} are extensively used to generate different configurations, stress the platform and analyze its performance.

\subsubsection{Pure Mininet}

Right from the first experiments it is possible to notice an irregular behavior at beginning and end of simulation. This happens because Mininet takes some time to setup the network at startup, and similarly at the end when the components start to be cleaned. This behavior can be expected and we took the straightforward approach to solving it, which is to neglect the first and last 10\% of simulation when analyzing the data.

\begin{figure}[ht]
\centering
\includegraphics[width=1\textwidth]{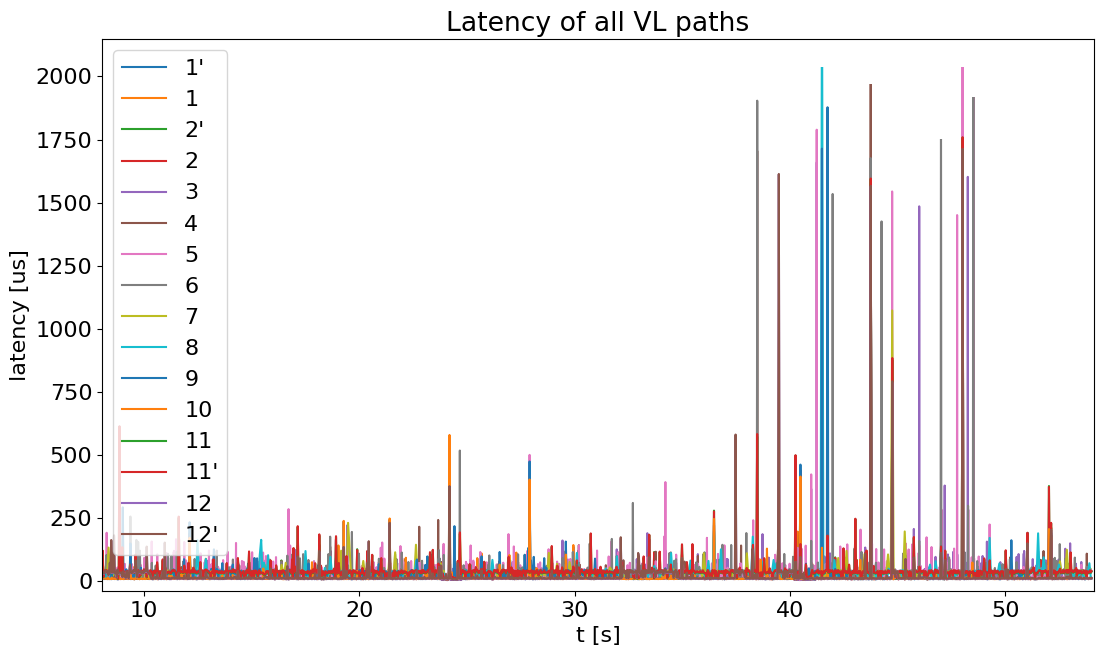}
\caption{\label{fig:puremininet1}Latency of all VL paths.}
\end{figure}

Figure \ref{fig:puremininet1} shows the latency of all VLs of the use case network topology obtained from a 60-seconds simulation. The x-axis in this plot represents the timestamp during simulation when a frame has been emitted by a sender process (in seconds), while the y-axis shows the network traversal time that each respective frame took to reach its destination (in microseconds), for all configured paths, calculated by the latency monitor.

\begin{figure}[ht]
\centering
\includegraphics[width=1\textwidth]{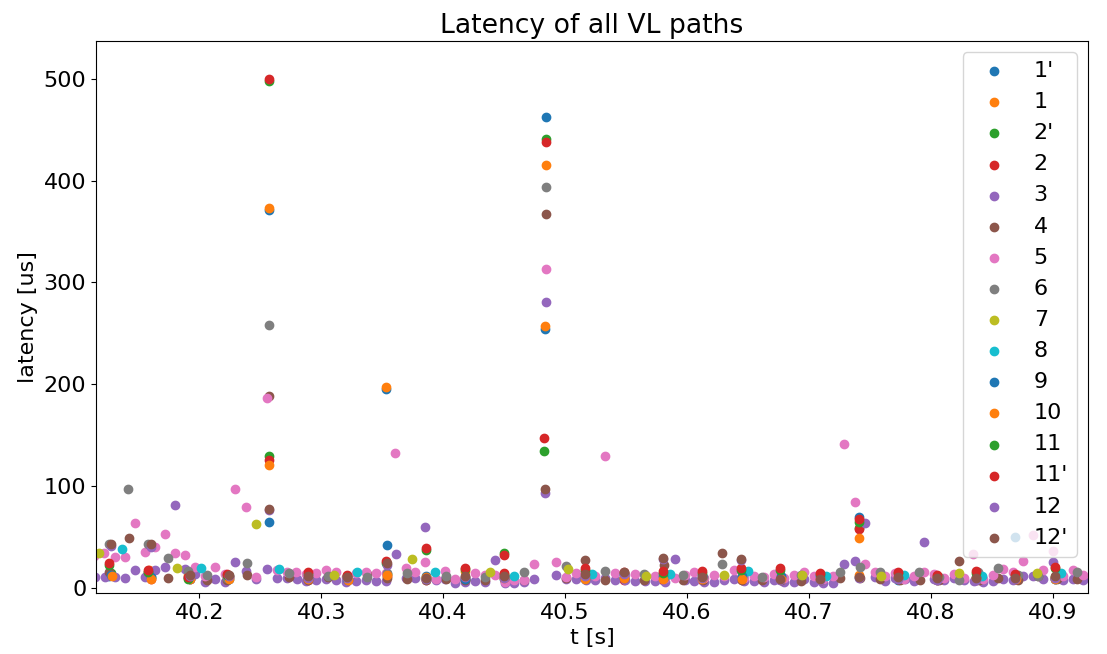}
\caption{\label{fig:puremininet2}Latency peaks in several VL paths.}
\end{figure}

In all simulations there are significant peaks in latency. Upon finer inspection (e.g. Figure \ref{fig:puremininet2}) we can see that it occurs with several VLs in an orderly fashion. One plausible explanation is that other services running on the OS are stalling the execution of the processes and releasing them at virtually the same time, which gives this flush-like characteristic. Since these experiments are done in a Virtual Machine, there are aspects of the underlying system that are not well-defined, and we assume that this kind of behaviour will disappear once we run on a laptop running a native Linux environment.

\begin{figure}[ht]
\centering
\includegraphics[width=1\textwidth]{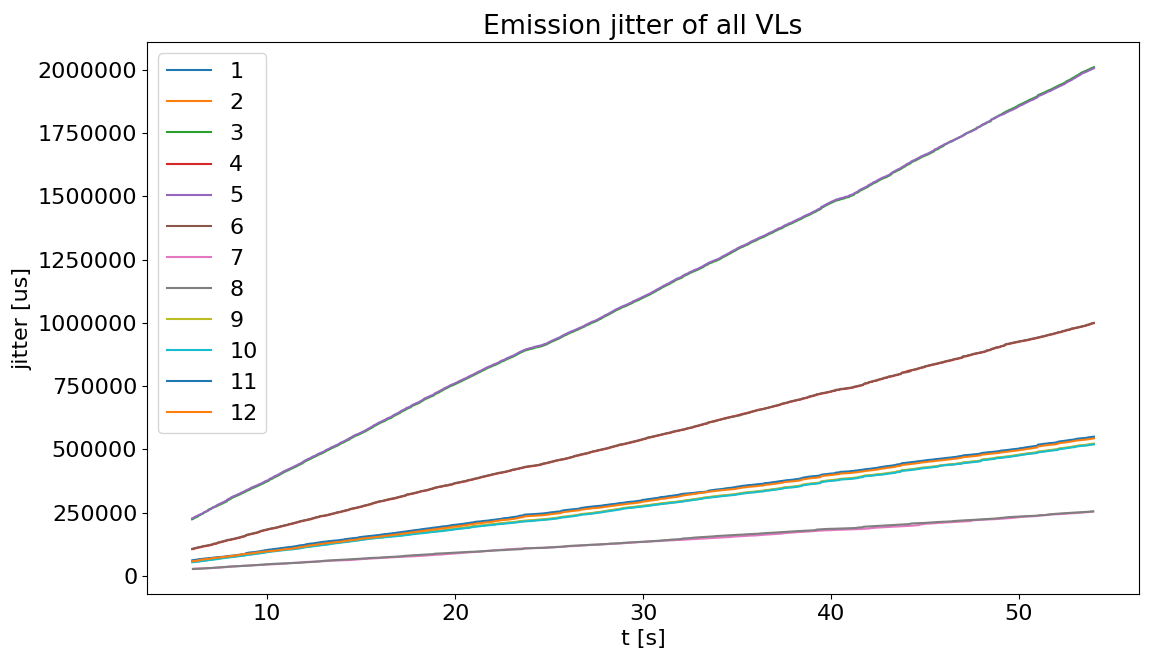}
\caption{\label{fig:puremininet3}Emission jitter accumulating on all VLs.}
\end{figure}

Another issue appeared when analyzing the jitter at frame emission, where it becomes evident that the periodicity dictated by the BAG is not being respected. Figure \ref{fig:puremininet3} illustrates the emission jitter of all VLs being accumulated as time advances. This accumulation of jitter is due to the way the periodic behavior of sender hosts is implemented. As shown in Listing \ref{lst:send}, the emission period is simulated by freezing execution for BAG $ms$ with \texttt{nanosleep()} before sending the next frame. As the tasks leading to the next sleep cycle are not instantaneous, the delay introduced in each period accumulates throughout execution. To counteract this unwanted behavior, we make use of the real-time Xenomai framework to implement the sender processes as a real-time periodic task.

\subsubsection{Xenomai Periodic Task}

Listing \ref{lst:xenomai} shows the code necessary to integrate the
real-time capabilities of Xenomai into the sender host nodes in order
to eliminate the accumulating jitter produced by using
\texttt{nanosleep()} to simulate periodic behavior.

Xenomai libraries are included by the \texttt{alchemy/task.h} and
the \texttt{alchemy/timer.h} statements, which contains the required
resources from Xenomai's Alchemy API to implement a real-time periodic
task.

The \texttt{mlockall()} function is actually a function provided by
Linux rather than Xenomai and is provided by the <sys/mman.h>
include. As real-time tasks can miss its deadlines if the task is
swapped out of memory by the operating system, this function call
makes sure that the memory that is currently mapped to the address
space of the process, as well as any memory that gets mapped into it
in the future, is “locked” into RAM and cannot get swapped out.

The \texttt{rt\_task\_shadow()} turns the current Linux task into a
real-time task using Xenomai's Alchemy API. The third parameter is the
\textit{priority} of the task, which tells the real-time scheduler how
important the task is. Higher priority tasks can interrupt lower
priority tasks, with 99 being the highest priority. The full
documentation can be found on the Xenomai website
\footnote{http://www.xenomai.org}.

Once \texttt{rt\_task\_shadow()} is called, the real-time task starts
executing. To make a Xenomai task periodic, we need to call the
\texttt{rt\_task\_set\_periodic()} function. \texttt{TM\_NOW} tells
Xenomai to start timing the task right away and PERIOD is the period
of the task in ticks of the clock. Since the default resolution of the
clock is 1 nanosecond, this argument is the same as the period you
want for the task expressed in nanoseconds.

In the \texttt{for} loop, instead of simulating the periodic behavior
by sleeping with \texttt{nanosleep()}, we use the
\texttt{rt\_task\_wait\_period()}, which blocks the loop till the
start of the next release point defined in the processor time line in
respect to the specified period.

\begin{lstlisting}[language=C,float=bt,captionpos=b,caption=\protect{Example code to add real-time periodic task capabilities to task running at sender host nodes.},label=lst:xenomai,frame=single]
#include <alchemy/task.h>
#include <alchemy/timer.h>
#include <sys/mman.h>

/* Configure socket */
/* Build AFDX frame */

mlockall(MCL_CURRENT | MCL_FUTURE);     // lock memory to avoid memory swapping
rt_task_shadow(NULL, "task", 99, 0);    // turns caller into real-time task

rt_task_set_periodic(NULL, TM_NOW, PERIOD);  // start task

for (int i = 0; i < COUNT; i++) {       // send COUNT frames periodically
    ...
    rt_task_wait_period(NULL);          // wait for next period
}

/* Save timestamps to file */
\end{lstlisting}

\begin{figure}[ht]
\centering
\includegraphics[width=1\textwidth]{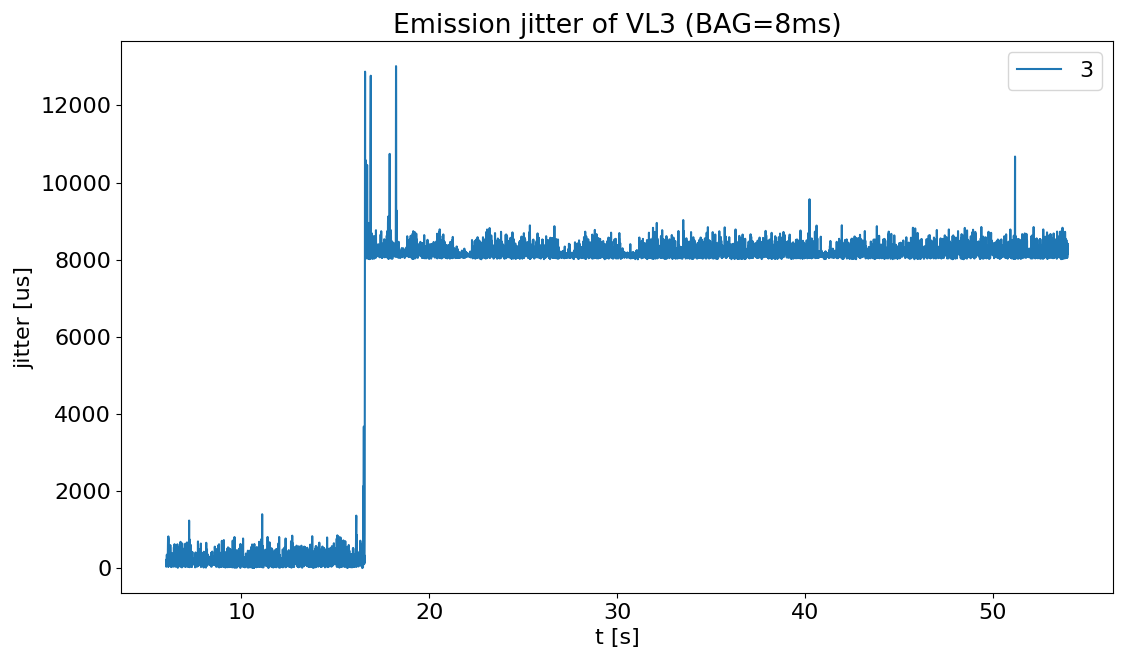}
\caption{\label{fig:periodictask}Emission jitter of $VL_{3}$. The 8\,000 microseconds step corresponds to one skipped period by the periodic task.}
\end{figure}

Figure \ref{fig:periodictask} shows the jitter of $VL_{}$ with the new implementation running with Xenomai dependencies. Now the jitter is not accumulating anymore, instead it is oscillating around a fixed value with mean close to zero, as expected from a periodic behavior. However, on smaller BAGs such as 8$ms$, there are steps on the measured jitter indicating the \emph{skipped periods} described on the previous section. This happens when the periodic task is not able to respect the release points, resulting on the processes skipping periods---also called \emph{period overruns} on the Xenomai documentation.

As much as skipping periods is undesirable, it is not considered unacceptable because frames can get lost on the network by other means (e.g. traffic policing). By implementing a mechanism on the jitter monitor, we can calculate the number of skipped periods to increment the fr`ame's sequence number accordingly, thus considering frames not sent in their respective periods as \emph{lost in emission}. Furthermore, in a more dedicated simulation setup---a laptop running a native Linux environment, for instance---this only happens at much smaller BAGs since the processes have substantially more resources to respect the release points and meet the deadlines.

Besides getting rid of the jitter accumulation, the latency peaks discussed previously did not disappear on the experiments running with Xenomai. More so, these peaks often coincide with jitter peaks, which do not have an explicit influence on the network traversal time. This correlation reinforce the assumption that the problem is with the scheduling, clock or resource management of the Virtual Machine.

\subsubsection{Native Linux Environment}

To get out of the constraints inherent of using a Virtual Machine, I configured a dedicated laptop to conduct the simulations. The following experiments were done on a standard Linux PC (Intel Core i7, 16GB Memory).

\begin{figure}%
    \centering
    \subfloat[Latency CDF]{{\includegraphics[width=1\textwidth]{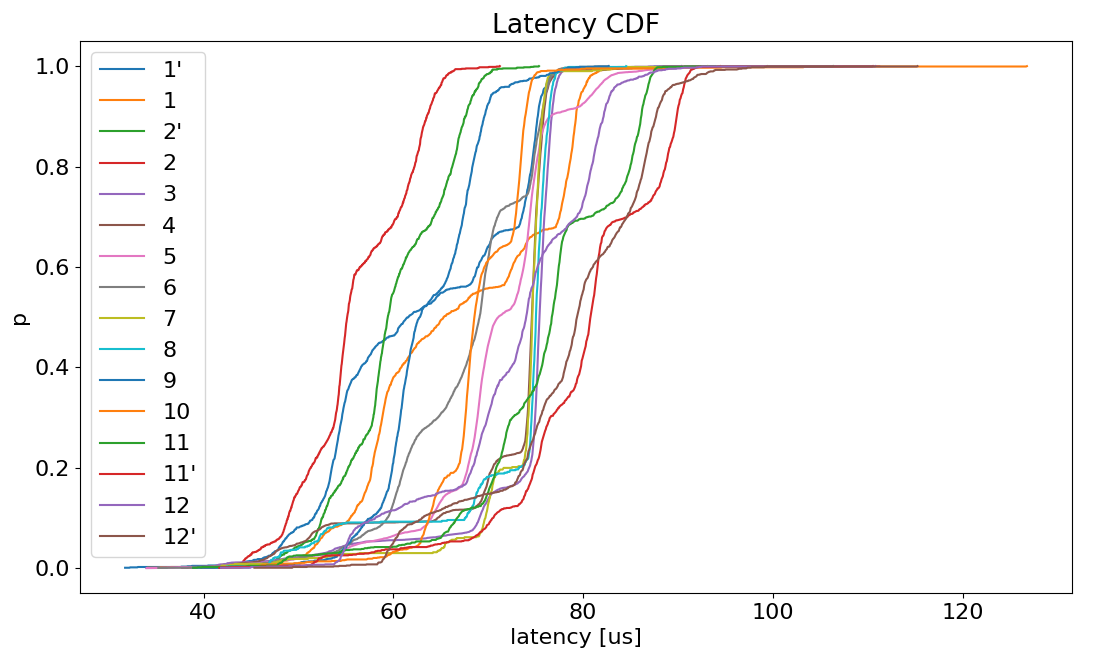} }}%
    \qquad
    \subfloat[Jitter CDF]{{\includegraphics[width=1\textwidth]{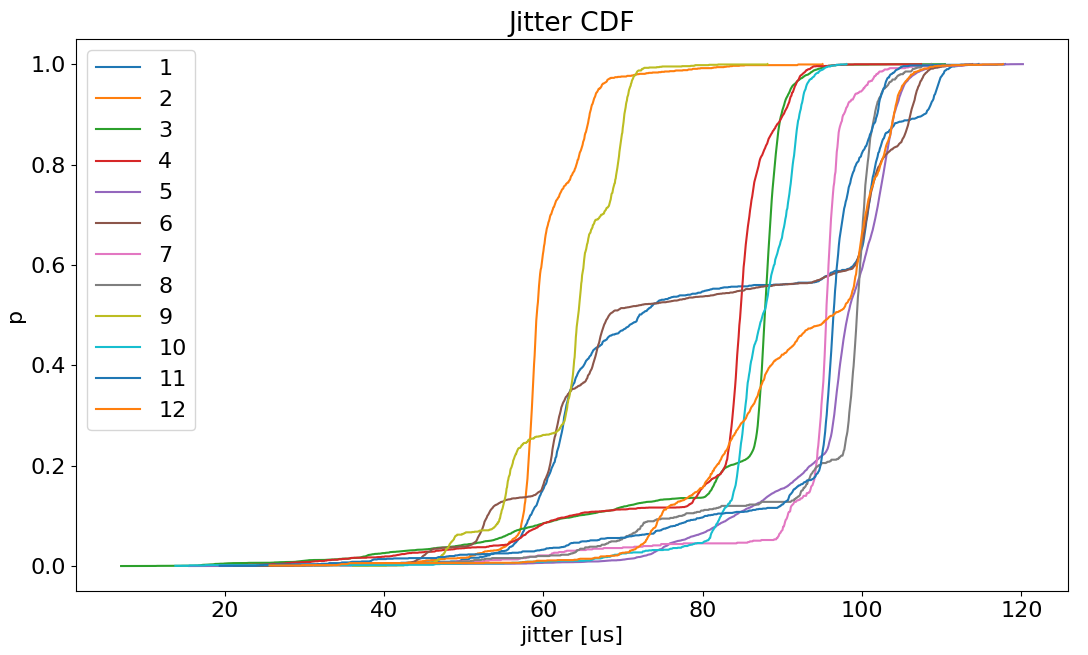} }}%
    \caption{CDF of latency and jitter of the use case network.}%
    \label{fig:latencyjitter}%
\end{figure}

Since the previous described abnormalities are not evident on the time-series of the new results, a better analysis can be done using the Cumulative Distribution Function (CDF), explained in Section~\ref{sec:benchmarking}. Figure \ref{fig:latencyjitter} show the CDF for the jitter and latency of a 60-seconds simulation of the use case network.

On the latency CDF it is still evident the presence of peaks, shown by this kind of plot by the straight lines on the upper right. Since they are rare, sparse occurrences (i.e. outliers), the probability of latencies being smaller than these values is almost 100\%. However, when performing the same analysis on the time-series as featured previously, I observed that the occurrences do not have the same profile as observed on simulations from VMs.

Emission jitter also improved considerably. Where before it oscillated around 500 $\mu$s for this same configuration, the maximum observed in this simulation is 120 $\mu$s, for instance. In addition to the faster periodic execution, period overruns only start to happen with much smaller periods (< 1 ms) and heavier loads. Nonetheless, even if these values are apparently small, they can still be considered over the limit in respect to AFDX standards. Considering we do not multiplex multiple VLs into a single connection, where jitter would most likely increase due to the traffic shaping mechanism, the measured jitter can be viewed as purely \emph{technological jitter}, which has a maximum admissible value of 40 $\mu$s according to the specification (see Sect.~\ref{sec:jitter}). Here the limitations of pure virtualization plays a part on underachieving the small jitters found on real applications. 

Primarily, all the processes running periodic tasks are sharing the resources of a single machine, in contrast with the distributed nature of Integrated Modular Avionics systems---such as AFDX. An implementation making better use of multi-threading can improve greatly the jitter, but to arrive at the same level as dedicated architectures is most likely impractical. 

Secondly, hard RTOSes like Xenomai guarantee the meeting of deadlines, but no assurance that it will do so as quickly as possible. This results in frames being sent much after the VL periodic release, but before the next period (\emph{too late} frames, as specified on Sect.~\ref{sec:jitter-monitor}). Another collateral effect of this behavior is frames being dropped by the traffic policing mechanism; in the event of the succeeding frame respecting the period more strictly, the time gap between the two consecutive frame shortens, making the instant bandwidth higher than the \emph{rate} configured on the token bucket algorithm.

On bigger, more complex network configurations, there are still a number of unexplained issues. Perhaps the most blatant one is the presence of negative jitters. To have negative jitter means that the periodic tasks are executing before its release point---defined by Xenomai---on the processor timeline, which is not supposed to happen. However, the existence of such jitters on data collected from simulations is most likely due to the timestamping process instead of faulty periodic behavior, which is very costly and sometimes uncertain.


\section{Conclusions and Future Work}

One of the main objective of this work is to evaluate the possible
benefits of using formal models and virtualization technologies for
the simulation of communication networks. While we concentrate on AFDX
networks here, the same approach could be used with other protocols.

This work had several requirements. Some of these requirements were
technical, such as developing a new virtualization framework for AFDX
based on the Mininet network emulator. And some were theoretical, such
as defining new formal models of AFDX networks that could precisely
take into account the temporal constraints of a network. This gave me
the opportunity to work with many different technologies: formal
verification; networking and virtualization; programming for real-time
OS; etc. which was a very rewarding experience.

During this project we developed three formal specifications for an
AFDX network with different levels of approximation---(i) timed
channels; (ii) direct virtual links; (iii) switched virtual
links---each one increasing the precision of how frames are treated
and transmitted through the network. I have also developed and
enriched a network virtualization framework, integrating real-time
features and implementing monitors to capture information about
latency and jitter at runtime.

The framework developed with Mininet allows the virtualization of an
AFDX network on a standard Linux computer, (we simply need to update
the kernel in order to support the real-time capabilities added by
Xenomai). This framework provides a local environment for testing
network configurations, is easy to setup, and scalability mainly
depends on the physical platform. While it still doesn't have
sufficient guarantees to respect the strict timing constraints, our
initial benchmarks support our conviction that virtualization can be
used to run simulations faster than with a real network and with an
acceptable time-accuracy.

Most of the anomalies faced during development disappeared once the
environment was moved out of a Virtual Machine environment. From the
experiences done on a dedicated laptop it appears that the system is
able to simulate an AFDX network of moderate complexity with minor
drawbacks. Further work is necessary to use the system resources more
efficiently in order to decrease the jitter, improve the time-stamping
precision and reduce the overhead caused by the monitors, as well as
using the developed benchmarking tools to better profile the
virtualization performance.

An important result missing from this project is the transformation of
the formal specification into runtime monitors. When porting the
models described using the Fiacre language to run on the Hippo
execution engine, a fundamental flaw was discovered---related to how
periodicity is implemented by the engine---which hindered further
development. However, the translation process demonstrated to be
possible and can be easily done once the issue is resolved.

What can also be considered a major contribution of this work is the
definition of a formal model for the token bucket traffic policing
mechanism. When developing the virtualization framework there were
limitations and particularities on existing algorithms---specifically
Linux traffic control---that strayed from the behavior described on
the AFDX specification. With this model we have a finer control over
the mechanism operation to match exactly how it is done on a real
system.


\clearpage

\newpage

\bibliographystyle{plain}
\bibliography{bibli}

\newpage
\appendix 
\section{Timed Channels} \label{a1}
  
\clearpage
  
\begin{figure}
\centering
\includegraphics[width=0.7\textwidth]{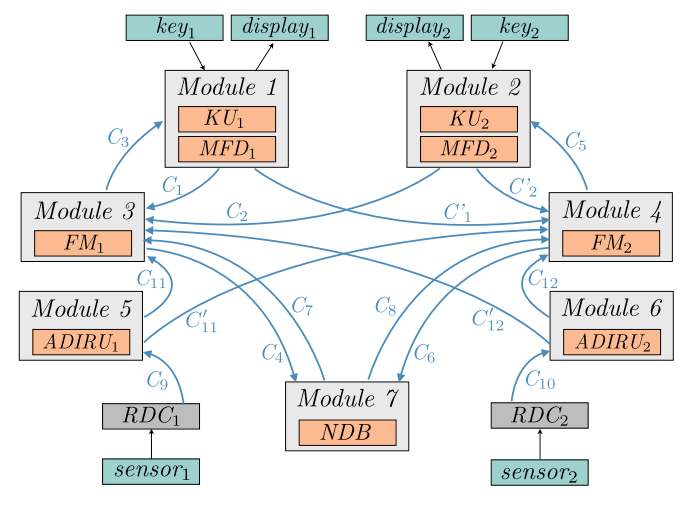}
\caption{\label{fig:fms-tc}Use case network architecture with timed channel approximation. Source: Lauer \cite{lauer}}
\end{figure}

\begin{table}
\centering
\caption{Timed channels parameters from the use case network}
\label{table:tc}
\begin{tabular}{@{}lllll@{}}
\toprule
channel   & source      & destination & BCTT (us) & WCTT (us) \\ \midrule
$C_{1}$   & $KU_{1}$    & $FM_{1}$    & 298       & 444       \\
$C_{1}'$  & $KU_{1}$    & $FM_{2}$    & 298       & 444       \\
$C_{2}$   & $KU_{2}$    & $FM_{1}$    & 298       & 444       \\
$C_{2}'$  & $KU_{2}$    & $FM_{2}$    & 298       & 444       \\
$C_{3}$   & $FM_{1}$    & $MFD_{1}$   & 310       & 490       \\
$C_{4}$   & $FM_{1}$    & $NDB$       & 310       & 450       \\
$C_{5}$   & $FM_{2}$    & $MFD_{2}$   & 310       & 490       \\
$C_{6}$   & $FM_{2}$    & $NDB$       & 310       & 450       \\
$C_{7}$   & $NDB$       & $FM_{1}$    & 400       & 508       \\
$C_{8}$   & $NDB$       & $FM_{2}$    & 400       & 508       \\
$C_{9}$   & $RDC_{1}$   & $ADIRU_{1}$ & 150       & 156       \\
$C_{10}$  & $RDC_{2}$   & $ADIRU_{2}$ & 150       & 156       \\
$C_{11}$  & $ADIRU_{1}$ & $FM_{1}$    & 452       & 584       \\
$C_{11}'$ & $ADIRU_{1}$ & $FM_{2}$    & 452       & 584       \\
$C_{12}$  & $ADIRU_{2}$ & $FM_{2}$    & 452       & 584       \\
$C_{12}'$ & $ADIRU_{2}$ & $FM_{1}$    & 452       & 584       \\ \bottomrule
\end{tabular}
\end{table}

\begin{lstlisting}[float=ht,captionpos=b,caption=\protect{Use case topology described as a CSV file to be loaded by the generator.},label=lst:csv-file,frame=single]
vlid,src,dst,bag,size
1,1,"3,4",32,75
2,2,"3,4",32,75
3,3,"1",8,625
4,3,"7",16,125
5,4,"2",8,625
6,4,"7",16,125
7,7,"3",64,500
8,7,"4",64,500
9,8,"5",32,64
10,9,"6",32,64
11,5,"3,4",32,87.5
12,6,"3,4",32,87.5
\end{lstlisting}


\end{document}